\documentclass[%
11pt,
twocolumn,
reprint,
superscriptaddress,
prd,
nofootinbib,
preprintnumbers,
]{revtex4-1}
\pdfoutput=1

\usepackage[utf8]{inputenc}
\usepackage{graphicx}
\usepackage{amsmath}
\usepackage{amssymb}

\usepackage{ascmac}
\usepackage[breaklinks=true]{hyperref}

\hypersetup{colorlinks,citecolor=TokiwaIro,linkcolor=RuriIro,urlcolor=RuriIro,linktocpage}

\usepackage{braket}
\usepackage{bm}

\usepackage[svgnames,table]{xcolor}
\usepackage{hhline}
\usepackage{multirow}
\usepackage{indentfirst}
\usepackage{latexsym}
\usepackage{mathtools}
\usepackage{cancel}
\usepackage{slashed}
\usepackage{colortbl}
\usepackage[mathscr]{euscript}
\usepackage{enumerate}
\usepackage{subcaption}
\usepackage{comment}
\usepackage{amsthm}

\allowdisplaybreaks[4]

\usepackage{tikz}
\usetikzlibrary{positioning,intersections,calc,arrows.meta}

\definecolor{RuriIro}{rgb}{0.,0.28,0.60}
\definecolor{TokiwaIro}{rgb}{0.,0.39,0.16}
\definecolor{AkaneIro}{rgb}{0.72,0.16,0.18}

\definecolor{dred}{rgb}{0.7,0.15,0.09}
\definecolor{dblue}{rgb}{0,0.12,0.64}
\definecolor{dgreen}{rgb}{0.2,0.51,0.19}
\definecolor{pegn}{rgb}{0.33,0.51,0.14}

\definecolor{kblue}{rgb}{0,0.48,0.73}
\definecolor{kred}{rgb}{0.73,0.25,0}
\definecolor{kgreen}{rgb}{0.48,0.73,0}

\definecolor{rgreen}{HTML}{7BAA17}
\definecolor{rred}{HTML}{AB1732}
\definecolor{rblue}{HTML}{007BBB}

\theoremstyle{plain}

\theoremstyle{remark}

\newcommand{\nn}{\nonumber}

\newcommand{\bb}{\mathbb}

\begin{document}

\title{
Non-Abelian $A_4$ vortices in  $SO(3)$
gauge theory
and non-invertible symmetries
}

\author{Yoshihiko Abe}\thanks{yabe3@keio.jp}
\affiliation{Graduate School of Science and Technology, Keio University, Yokohama, Kanagawa 223-8522, Japan}
\affiliation{Keio University Sustainable Quantum Artificial Intelligence Center (KSQAIC), Keio University, Tokyo 108-8345, Japan}
\affiliation{Quantum Computing Center, Keio University, 3-14-1 Hiyoshi, Kohoku-ku, Yokohama, Kanagawa 223-8522, Japan}

\author{Tetsutaro Higaki}\thanks{thigaki@rk.phys.keio.ac.jp}
\affiliation{Department of Physics, Keio University, Yokohama 223-8533, Japan}
\affiliation{Reserach and Education Center for Natural Sciences, Keio University, 4-1-1 Hiyoshi, Yokohama, Kanagawa 223-8521, Japan}

\author{Kazuya Murakami}\thanks{kazuyamurakami@keio.jp}
\affiliation{Department of Physics, Keio University, Yokohama 223-8533, Japan}

\author{Muneto Nitta}\thanks{mune.nitta@gmail.com}
\affiliation{Department of Physics, Keio University, 4-1-1 Hiyoshi, Yokohama, Kanagawa 223-8521, Japan}
\affiliation{Reserach and Education Center for Natural Sciences, Keio University, 4-1-1 Hiyoshi, Yokohama, Kanagawa 223-8521, Japan}
\affiliation{International Institute for Sustainability with Knotted Chiral Meta Matter (WPI-SKCM$^2$), Hiroshima University, 1-3-2 Kagamiyama, Higashi-Hiroshima, Hiroshima 739-8531, Japan}

\author{Ryo Yokokura}\thanks{ryokokur@keio.jp}
\affiliation{Department of Physics, Keio University, 4-1-1 Hiyoshi, Yokohama, Kanagawa 223-8521, Japan}
\affiliation{Reserach and Education Center for Natural Sciences, Keio University, 4-1-1 Hiyoshi, Yokohama, Kanagawa 223-8521, Japan}

\begin{abstract}
We construct finite-tension non-Abelian vortex solutions in a renormalizable $(3+1)$-dimensional $SO(3)$ gauge theory Higgsed to the tetrahedral group $A_4$ by a Higgs field in the spin-3 representation. Since the vacuum manifold is $SO(3)/A_4$, the vortices are characterized by the non-Abelian fundamental group $\pi_1(SO(3)/A_4)\simeq \widetilde{A}_4$, the binary tetrahedral group. We obtain explicit axisymmetric vortex solutions carrying holonomies corresponding to the order-two and order-three conjugacy classes of $A_4$, determine their tensions numerically, and show that they exhibit type-I, type-II, and Bogomol'nyi--Prasad--Sommerfield-like behavior depending on the Higgs and gauge boson mass ratios. The vortices are classified by conjugacy classes of $\widetilde{A}_4$, while their infrared descriptions are labeled by conjugacy classes of $A_4$. We further demonstrate that the smooth finite-tension vortices reduce in the infrared to Gukov--Witten surface operators of the $A_4$ discrete gauge theory, thereby establishing a finite-energy ultraviolet completion of non-invertible defects in a renormalizable gauge-Higgs theory.

\end{abstract}

\maketitle

\section{Introduction and summary}

Topological solitons, which are classical solutions with topological
charges, have been extensively studied in past decades~\cite{Vilenkin:2000jqa,Manton:2004tk,Weinberg:2012pjx}.
Among others, vortices can often arise in many contexts such as
superconductors, superfluids, cosmic strings, and so on \cite{Abrikosov:1956sx,Nielsen:1973cs}.
Here, they are codimension-2 objects whose topological charges are
winding numbers.
It has been known that a non-Abelian fundamental group
$\pi_1 (G/H)$ of a non-Abelian symmetry group $G$ and its unbroken subgroup $H$ allows {\it non-Abelian vortices}
exhibiting exotic properties
such as non-Abelian fusion and braiding rules~\cite{Bais:1980vd,Alford:1990mk,Wilczek:1989kn,Alford:1990ur,Alford:1992yx,Bucher:1991bc,Lo:1993hp,Brekke:1992he,Brekke:1997jj,McGraw:1997nx}.\footnote{
Note that the terminology ``non-Abelian vortices'' implies another meaning in other contexts
 such as
supersymmetric QCD
\cite{Hanany:2003hp,Auzzi:2003fs,Eto:2005yh,Eto:2006cx,Eto:2006pg,Shifman:2009zz}
and dense QCD \cite{Balachandran:2005ev,Nakano:2007dr,Eto:2009kg,Eto:2009bh,Eto:2009tr,Eto:2013hoa}, where even though $\pi_1 (G/H)$ is Abelian,
vortices are called non-Abelian in the sense that
$H$ is non-Abelian
and they carry non-Abelian moduli.
}
Global analogues of such non-Abelian vortices
were first found in condensed matter physics
\cite{Mermin:1979zz}:
liquid crystals~\cite{Poenaru1977,vol77,Mermin:1979zz,Lavrentovich2001},
$^3$He superfluids~\cite{Balachandran:1983pf,salomaaRMP,Volovik:2003fe},
and Bose-Einstein condensates (BECs)
\cite{Semenoff:2006vv,Kobayashi:2008pk,Kawaguchi:2012ii,Borgh:2016cco,Mawson:2018klj,Annala:2022bdd,Rajamaki:2023ymv,Kobayashi:2024aip},
as well as in nuclear and astrophysics such as
$^3P_2$ neutron superfluids in neutron stars
\cite{Masuda:2016vak,Masaki:2021hmk,Masaki:2023rtn,Kobayashi:2022moc,Kobayashi:2022dae,Marmorini:2020zfp,Hattori:2026bvb}.
In $(3+1)$ dimensions,
the collision of two vortices corresponding to noncommutative $\pi_1$ elements results in
a rung vortex
\cite{Poenaru1977,Mermin:1979zz,Kobayashi:2008pk},
and consequently, in cosmology, it yields
a non-Abelian cosmic string network \cite{McGraw:1997nx}.
In addition, non-Abelian vortex knots are also possible
\cite{Brekke:1992he,Annala:2022bdd,Rajamaki:2023ymv,Kobayashi:2024aip}.

In gauge theories with gauge group $G$ Higgsed
to $H$,
non-Abelian vortices exhibit non-Abelian Aharonov--Bohm (AB) effects if $\pi_1 (G/H)$ is non-Abelian~\cite{Horvathy:1985jr,Sundrum:1986ub,Alford:1989ch,Preskill:1990bm,Bais:1991pe,Bais:1992ca}.
The low-energy limit of the Higgs phase can be described by
an $H$ gauge theory
where there are no local fluctuations~\cite{Krauss:1988zc}.
The non-Abelian fusion and braiding rules of the vortices
in the gauge theories have been discussed in terms of the
AB effects for charged particles far from the vortices.
However, to the best of our knowledge, explicit vortex solutions with non-Abelian \(\pi_1(G/H)\) have not been constructed so far.
Thus, several important physical properties, such as their tensions and the interactions between vortices, have remained unclear.

In this article, we derive the vortex solutions
with non-Abelian $\pi_1 (G/H)$
in a Higgs phase of a $(3+1)$-dimensional gauge theory.
We focus on an $SO(3)$ gauge theory Higgsed to
the tetrahedral group $A_4$, where $\pi_1 (SO(3)/A_4)$ is non-Abelian.
The discrete group $A_4$ has been widely used in particle-physics flavor models and, more recently, in models based on modular symmetry~\cite{Altarelli:2010gt,Feruglio:2017spp,Ishimori:2010au}.
$SO(3)$ gauge theories Higgsed to $A_4$ have also been studied in particle physics~\cite{Ovrut:1977cn,Etesi:1996urw,King:2018fke}
\footnote{
In condensed matter physics,
similar $A_4$ vortices exist in
the so-called cyclic phase
of spin-2 BECs in which
$U(1) \times SO(3)$ is broken to a subgroup containing $A_4$ (often denoted by $T$ in that context)
\cite{Semenoff:2006vv,Kobayashi:2008pk,Mawson:2018klj,Kobayashi:2024aip,Kawaguchi:2012ii}.
}.
We show that the vortex solutions can be classified by
the conjugacy class of the unHiggsed subgroup $A_4$,
and the non-Abelian AB effects can be expressed by
the indicator of $A_4$.
This is the first explicit finite-tension realization of non-Abelian vortices whose non-Abelianity originates solely from a non-Abelian fundamental group.

Using the field configurations of the vortices, we discuss
the tensions of the vortices.
The tensions depend on the ratio of the masses of
the Higgs field and the gauge field.
The dependence implies that the vortices are attractive
for a large gauge field mass
and repulsive for a large Higgs field mass.
We find that there are no attractive or repulsive forces if
the mass of the gauge field is equal to that of the Higgs field.
In this case,
the vortices might be understood as Bogomol'nyi--Prasad--Sommerfield (BPS)
states.
See e.~g.~refs.~\cite{Eto:2006pg,Shifman:2009zz}
as a review for BPS vortices.
However, our vortices are
quite different from conventional BPS states because
the mass-charge relations do not hold due to $\pi_1 (G/H) \not\supset \bb{Z}$.

Finally, we discuss our vortex solutions from the viewpoint of
generalized global symmetries~\cite{Gaiotto:2014kfa}.
It has been known that the $A_4$ gauge theory possesses
a non-invertible one-form symmetry, whose symmetry generators are the so-called Gukov--Witten operators,
which produce non-Abelian AB effects for test particles~\cite{Gukov:2008sn,Heidenreich:2021xpr}.
We find that the non-Abelian vortices exhibit the same fusion rule and
the non-Abelian AB effects as the Gukov--Witten operators.
This coincidence means that
the Gukov--Witten operators can be realized
as the low-energy limit of the non-Abelian vortices.
Since our gauge-Higgs system is a renormalizable UV completion of the infrared \(A_4\) gauge theory,
the non-invertible one-form symmetry, which emerges only in the Higgs phase,
does not survive once the gauge group \(SO(3)\) is unHiggsed, for example, at sufficiently high energies.
In the infrared \(A_4\) gauge theory, the non-invertible one-form symmetry is explicitly broken in the presence of dynamical matter fields.

This article is organized as follows.
In section \ref{sec:model},
we introduce our model.
Using the model, the vortex solutions are
established in section \ref{sec:solutions}.
We then calculate the tensions of the vortices,
and find type-I, -II, and BPS-like vortices in section \ref{sec:type_I_II_BPS}.
The topological classification of the vortices is discussed in
section \ref{sec:topological_classification}.
In section \ref{sec:noninvertible_symmetry},
we relate our vortex solutions to the non-invertible symmetries.
Finally, we discuss our results in section \ref{sec:discussion}.
In Appendix~\ref{app:spin3_generators}, we give the explicit \(7\times7\) matrices representing the Lie algebra in the spin-3 representation.
In Appendix~\ref{app:quartic_invariant}, we write down our Higgs potential explicitly.
The uniqueness of the profile for a Higgs field needed in section \ref{sec:solutions} is
derived in Appendix~\ref{app:b_uniqueness_eta1}.

\section{Model}\label{sec:model}
We consider the following \(SO(3)\) gauge-Higgs system in \(3+1\) dimensions,
with Lagrangian density
\begin{align}
    \mathcal{L} = - \frac{1}{4e^2}G_{\mu\nu}^{a}G^{a\mu\nu} - \frac{1}{2} D_\mu\phi_A\,D^\mu\phi_A - V(\phi). \label{eq:lagrangian}
\end{align}
Here \(e\) is the gauge coupling, and \(\phi\) is a Higgs field in the spin-3 irreducible representation of \(SO(3)\).
We expand it as
\begin{align}
    \phi(x)=\sum_{A=1}^{7}\phi_A(x)e_A ,
\end{align}
where \(e_A\), \(A=1,\ldots,7\), is an orthonormal basis of the spin-3 representation space introduced below.

We write the \(SO(3)\) gauge field as
\begin{align}
    W_\mu = W_\mu^a T^a .
\end{align}
We take \(iT^a\) to be the following generators of the Lie algebra \(\mathfrak{so}(3)\):
\begin{align}
    iT^1 =
    \begin{pmatrix}
        0 & 0 & 0\\
        0 & 0 & 1\\
        0 &-1& 0
    \end{pmatrix},
    iT^2 =
    \begin{pmatrix}
        0 & 0 &-1\\
        0 & 0 & 0\\
        1& 0 & 0
    \end{pmatrix},
    iT^3 =
    \begin{pmatrix}
        0 & 1& 0\\
        -1& 0 & 0\\
        0 & 0 & 0
    \end{pmatrix}.
\end{align}
With this convention, the field-strength components are
\begin{align}
  G_{\mu\nu}^a = \partial_\mu W_\nu^a - \partial_\nu W_\mu^a - \varepsilon^{abc} W_\mu^b W_\nu^c, \label{eq:field_strength}
\end{align}
where \(\varepsilon^{abc}\) is the totally antisymmetric tensor with \(\varepsilon^{123}=1\).

A gauge transformation is given by an \(SO(3)\)-valued function \(U(x)\).
With our convention, it acts on the fields as
\begin{align}
    W_\mu &\longmapsto U W_\mu U^{-1} + i(\partial_\mu U) U^{-1},\\
    \phi &\longmapsto \rho(U)\phi ,
\end{align}
where \(\rho\) denotes the spin-3 representation of \(SO(3)\).

We now describe the spin-3 representation more explicitly.
We denote its representation space by \(\mathcal H_3\).
It can be realized as the space of real harmonic homogeneous polynomials of degree three in three variables,
\begin{align}
    \mathcal{H}_3 := \left\{f \in \mathbb{R}[\mathcal X,\mathcal Y,\mathcal Z] \,\middle|\, \deg f=3,\ \Delta f=0 \right\}.
\end{align}
The \(SO(3)\) action is given by
\begin{align}
    (\rho(g)f)\bigl((\mathcal X,\mathcal Y,\mathcal Z)^T\bigr)
    = f\bigl(g^{-1}(\mathcal X,\mathcal Y,\mathcal Z)^T\bigr),\,\, g \in SO(3).
\end{align}
With this action, the covariant derivative is
\begin{align}
    D_\mu\phi_A = \partial_\mu\phi_A + W_\mu^a (X_a)_{AB}\phi_B,
\end{align}
where \(X_a\) are the \(7\times 7\) matrices representing the Lie algebra in the spin-3 representation.
Their explicit form is given in Appendix~\ref{app:spin3_generators}.

We define the \(SO(3)\)-invariant inner product by
\begin{align}
    \langle f,g\rangle := \frac{1}{4\pi} \int_{S^2} f(\mathcal X,\mathcal Y,\mathcal Z)g(\mathcal X,\mathcal Y,\mathcal Z) \,d\Omega, \label{eq:H3_inner_product}
\end{align}
where \(S^2\subset\mathbb R^3\) is the unit sphere,
\begin{align}
    S^2 = \left\{ (\mathcal X,\mathcal Y,\mathcal Z)\in\mathbb R^3 \,\middle|\, \mathcal X^2+\mathcal Y^2+\mathcal Z^2=1 \right\},
\end{align}
and \(d\Omega\) denotes its standard area element.
With respect to this inner product, we use the following orthonormal basis of \(\mathcal H_3\):
\begin{equation}
    \begin{split}
    e_1 &= \sqrt{7} \left(\mathcal{X}^3 - \frac{3}{2} \mathcal{X}(\mathcal{Y}^2+\mathcal{Z}^2)\right)\\
    e_2 &= \sqrt{7} \left(\mathcal{Y}^3 - \frac{3}{2} \mathcal{Y}(\mathcal{Z}^2+\mathcal{X}^2)\right)\\
    e_3 &= \sqrt{7} \left(\mathcal{Z}^3 - \frac{3}{2} \mathcal{Z}(\mathcal{X}^2+\mathcal{Y}^2)\right)\\
    e_4 &= \frac{\sqrt{105}}{2} \mathcal{X}(\mathcal{Y}^2-\mathcal{Z}^2)\\
    e_5 &= \frac{\sqrt{105}}{2} \mathcal{Y}(\mathcal{Z}^2-\mathcal{X}^2)\\
    e_6 &= \frac{\sqrt{105}}{2} \mathcal{Z}(\mathcal{X}^2-\mathcal{Y}^2)\\
    e_7 &= \sqrt{105} \mathcal{X}\mathcal{Y}\mathcal{Z}.
    \end{split}
\end{equation}
For \(\phi\in\mathcal H_3\), we define
\begin{align}
    I_n(\phi) := \frac{1}{4\pi} \int_{S^2} \phi(\mathcal{X},\mathcal{Y},\mathcal{Z})^n \,d\Omega.
\end{align}
In particular, if \(\phi=\sum_{A=1}^{7}\phi_A e_A\), then
\begin{align}
    I_2(\phi) = \langle \phi,\phi \rangle = \sum_{A=1}^{7}\phi_A^2 .
\end{align}
The \(SO(3)\)-invariant polynomials on \(\mathcal H_3\) of degree at most four are spanned by
\begin{align}
    1,\qquad I_2,\qquad I_2^2,\qquad I_4.
\end{align}
For the spin-3 representation, representation theory gives
\begin{align}
    \min_{\phi\in\mathcal H_3\setminus\{0\}} \frac{I_4(\phi)}{I_2(\phi)^2} = \frac{315}{143}.
\end{align}
We therefore normalize the quartic invariant as
\begin{align}
    Q(\phi) := \frac{143}{315}I_4(\phi) - I_2(\phi)^2.
\end{align}
With this normalization,
\begin{align}
    Q(\phi) \ge 0.
\end{align}
An explicit component expression for \(Q\) is given in Appendix~\ref{app:quartic_invariant}.
Using these invariants, we take the Higgs potential to be
\begin{align}
  V(\phi) &= \gamma Q(\phi) - \mu^2 I_2(\phi) + \lambda I_2(\phi)^2 + \frac{\mu^4}{4\lambda} \nonumber\\
  &= \gamma Q(\phi) + \lambda\left(I_2(\phi) - v^2\right)^2 ,
\end{align}
where
\begin{align}
  \gamma>0,\qquad \lambda>0,\qquad \mu^2>0, \qquad v^2=\frac{\mu^2}{2\lambda}.
\end{align}
With this choice, the vacuum expectation value of the Higgs field is
\begin{align}
 \left< \phi \right> = v e_7,
\end{align}
up to an \(SO(3)\) gauge transformation.
Since \(e_7\propto \mathcal{XYZ}\), only a finite subgroup of \(SO(3)\) leaves this vacuum invariant.
The remaining gauge symmetry is the tetrahedral group \(A_4\), generated for example by
\begin{align}
  r_2 =
  \begin{pmatrix}
      -1&0&0\\
      0&-1&0\\
      0&0&1
  \end{pmatrix},
  \qquad
  r_3 =
  \begin{pmatrix}
      0&0&1\\
      1&0&0\\
      0&1&0
  \end{pmatrix}.
  \label{eq:r2r3}
\end{align}
The first generator is a \(\pi\)-rotation around the \((0,0,1)\)-axis,
and the second is a \(2\pi/3\)-rotation around the \((1,1,1)\)-axis.
These generators satisfy
\begin{align}
r_2^2=r_3^3=(r_2r_3)^3=1_{3\times 3},
\end{align}
which are the defining relations of $A_4$.
Thus, the Higgs mechanism breaks the gauge symmetry as
\begin{align}
  SO(3)\to A_4 .
\end{align}
The corresponding vacuum manifold is
\begin{align}
  \mathcal M \simeq SO(3)/A_4.
\end{align}
The topological classification of finite-tension string configurations will be discussed in section \ref{sec:topological_classification}.
In the next section, we construct explicit axisymmetric vortex solutions with asymptotic holonomies \(r_2^n\) and \(r_3^m\) in \(A_4\).

\section{Vortex solutions}\label{sec:solutions}
We now construct static, \(z\)-independent, axisymmetric vortex solutions in the Higgs phase.
Let \((r,\theta,z)\) be cylindrical coordinates with the vortex aligned along the \(z\)-axis, where the vortex core is located at \(r=0\).
The finite-tension condition requires that, as \(r\to\infty\), the Higgs field approaches the vacuum manifold.
Thus, when one goes around a large circle \(C_\infty\) linking the vortex,
the Higgs field may return to itself up to an element of the unHiggsed group \(A_4\).
Equivalently, the asymptotic gauge holonomy
\begin{align}
  U_\infty=\mathrm{P}\exp\left(-i\oint_{C_\infty} W_\mu \,dx^\mu\right)
\end{align}
must belong to \(A_4\).
Physically, a nontrivial \(U_\infty\) represents magnetic flux carried along the vortex line.

Before presenting the vortex ansatze, we introduce the mass parameters and dimensionless variables used in the radial equations.
Expanding the Higgs potential around the vacuum \(\phi=ve_7\), the Hessian with respect to the orthonormal basis \(e_1,\ldots,e_7\) is
\begin{align}
  \left.\mathrm{Hess}\,V\right|_{ve_7} = \operatorname{diag}\left(m_-^2,\,m_-^2,\,m_-^2,\,0,\,0,\,0,\,m_+^2\right).
\end{align}
The three zero modes \(e_4,e_5,e_6\) are the would-be Nambu--Goldstone directions associated with the broken \(SO(3)\) generators and are eaten by the gauge field.
The two Higgs-sector mass scales are
\begin{align}
  m_+ = 2\mu, \qquad
  m_- = \frac{4}{3\sqrt{5}}\frac{\mu\sqrt{\gamma}}{\sqrt{\lambda}}.
\end{align}
The Higgs mechanism also gives a common mass to the three gauge bosons,
\begin{align}
  m_G = \frac{\sqrt{2}\mu e}{\sqrt{\lambda}}.
\end{align}
We define the dimensionless radial coordinate by
\begin{align}
  R = ev\,r,
\end{align}
and introduce the dimensionless mass ratios
\begin{align}
  \kappa := \frac{m_+}{m_G}, \qquad
  \eta := \frac{m_+}{m_-}.
  \label{eq:mass-ratio}
\end{align}

\subsection{Order-two vortex family}
We begin with a vortex family whose asymptotic holonomy is given by the \(N\)-th power of a fixed order-two element of \(A_4\).
For a positive integer \(N\), we take
\begin{align}
  W_\mu \,dx^\mu &= -\frac{N}{2} K(r) T^3 d\theta, \label{eq:ansatz1_gauge}\\
  \phi &= v h(r) \left(\sin N\theta\, e_6 + \cos N\theta\, e_7\right). \label{eq:ansatz1_higgs}
\end{align}
The finite-tension condition and continuity at the vortex core lead to the boundary conditions:
\begin{align}
  K(0) = 0, \qquad K(\infty) = 1,\\
  h(0) = 0, \qquad h(\infty) = 1 .
\end{align}
The holonomy around a large circle is
\begin{align}
  U_\infty &= \exp\left(N\pi iT^3\right)
  =
  \begin{pmatrix}
    -1&0&0\\
    0&-1&0\\
    0&0&1
  \end{pmatrix}^{N}.
  \label{eq:UN}
\end{align}
Thus, the \(SO(3)\) holonomy is the \(N\)-th power of an order-two element of \(A_4\).
We refer to this family as the order-two family.
The radial equations are obtained by substituting the ansatz into the Euler--Lagrange equations:
\begin{align}
	0 &= - K'' + \frac{K'}{R} + 4 h^{2} \left(K - 1\right),\\
  0 &= h'' + \frac{h'}{R} - \frac{N^{2} \left(K - 1\right)^{2} h}{R^{2}} + 2\kappa^2\left(1 - h^2\right) h.
\end{align}
Here and below, a prime denotes differentiation with respect to \(R\).

We now examine the asymptotic behavior of the profile functions.
First, linearizing the equations around the vacuum at spatial infinity, we find
\begin{align}
  K(r) &= 1 - C_G e^{-m_G r}+\cdots, \\
  h(r) &= 1 - C_+ e^{-m_+ r}+\cdots .
\end{align}
Near the origin, expanding the reduced equations around \(R=0\), one finds
\begin{align}
  K(R) = K_2R^2 + O(R^4),
  \qquad
  h(R) = h_NR^N + O(R^{N+2}) .
\end{align}
Here, \(C_G\), \(C_+\), \(K_2\), and \(h_N\) are constants.
This shows that the configuration is smooth in Cartesian coordinates near the core.

\begin{figure}[t]
  \centering
  \includegraphics[width=\columnwidth]{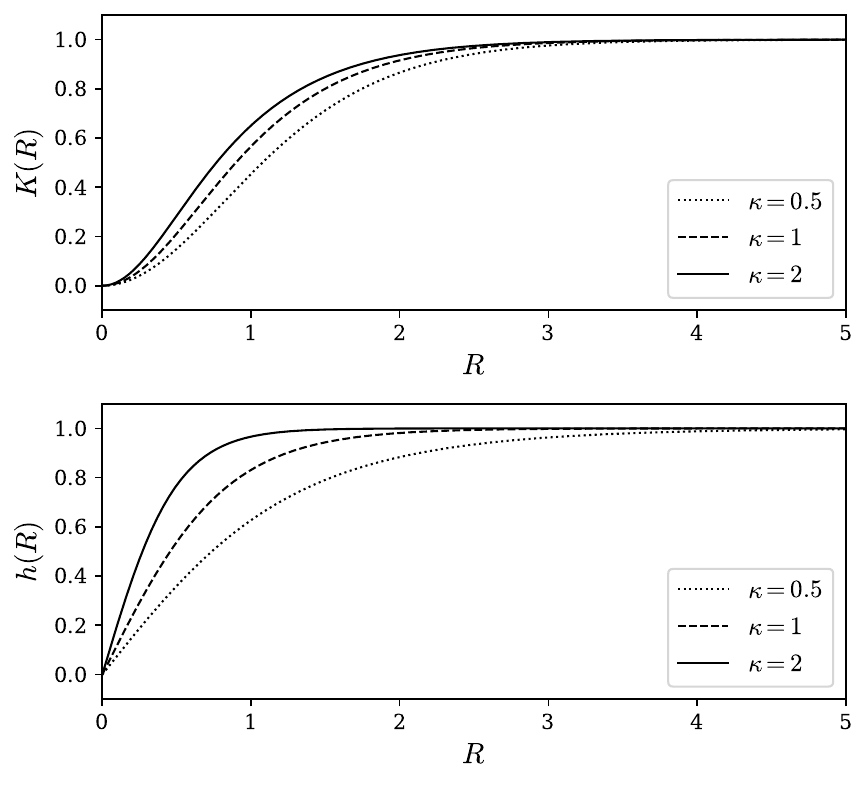}
  \caption{
    Numerical profiles \(K(R)\) and \(h(R)\) of the order-two vortex with \(N=1\) for \(\kappa=0.5,1,2\).
  \label{fig:order2_profiles}
  }
\end{figure}

Solving the above boundary value problem numerically, we obtain the representative profile shown in FIG.~\ref{fig:order2_profiles}.

\subsection{Order-three vortex family}
We next turn to a vortex family whose asymptotic holonomy is given by the \(M\)-th power of a fixed order-three element of \(A_4\).
Let
\begin{align}
  T^\Delta := T^1+T^2+T^3 .
\end{align}
For a positive integer \(M\), we take
\begin{align}
  W_\mu \,dx^\mu &= - \frac{M\sqrt{3}}{9}K(r)T^\Delta d\theta, \label{eq:ansatz2_gauge}\\
  \phi &= v \begin{pmatrix}
    \frac{2\sqrt{15}}{27}\left(a(r)\cos M\theta-b(r)\right)\\
    \frac{2\sqrt{15}}{27}\left(a(r)\cos M\theta-b(r)\right)\\
    \frac{2\sqrt{15}}{27}\left(a(r)\cos M\theta-b(r)\right)\\
    \frac{2\sqrt{3}}{9}a(r)\sin M\theta\\
    \frac{2\sqrt{3}}{9}a(r)\sin M\theta\\
    \frac{2\sqrt{3}}{9}a(r)\sin M\theta\\
    \frac{4}{9}a(r)\cos M\theta+\frac{5}{9}b(r)
  \end{pmatrix}. \label{eq:ansatz2_higgs}
\end{align}
The finite-tension condition and continuity at the vortex core lead to the boundary conditions:
\begin{align}
  K(0) = 0,\qquad K(\infty) = 1,\\
  a(0) = 0,\qquad a(\infty) = 1,\\
  b'(0) = 0,\qquad b(\infty) = 1 .
\end{align}
The holonomy around a large circle is
\begin{align}
  U_\infty &= \exp\left(\frac{2M\pi\sqrt{3}}{9} i(T^1+T^2+T^3)\right)
  = \begin{pmatrix}
    0&0&1\\
    1&0&0\\
    0&1&0
  \end{pmatrix}^{M}.
  \label{eq:UM}
\end{align}
Thus the \(SO(3)\) holonomy is the \(M\)-th power of an order-three element of \(A_4\).
We refer to this family as the order-three family.

Substituting this ansatz into the full Euler--Lagrange equations,  one finds
\begin{widetext}
    \begin{align}
      0 &= K'' - \frac{K'}{R} - 4 a^{2} \left(K - 1\right), \\
      0 &= - a'' - \frac{a'}{R} + \frac{M^{2} \left(K - 1\right)^{2} a}{R^{2}} + \frac{2\kappa^2\left(-9 + 4 a^{2} + 5 b^{2}\right) a}{9} + \frac{10\kappa^2 \left(a^2 - b^2\right) a}{9 \eta^{2}},
      \\
      0 &= - b'' - \frac{b'}{R} + \frac{2\kappa^2\left(-9 + 4 a^{2} + 5 b^{2}\right) b}{9} - \frac{8\kappa^2 \left(a^2 - b^2\right) b}{9 \eta^{2}}.
    \end{align}
\end{widetext}

We now examine the asymptotic behavior of the profile functions.
At spatial infinity, it is useful to introduce
\begin{align}
  h_+ := \frac{4a+5b}{9},
  \qquad
  h_- := a-b .
\end{align}
Linearizing the equations around the vacuum \(K=a=b=1\), we find
\begin{align}
  K(r) &= 1-C_G e^{-m_G r} + \cdots,\\
  h_+(r) &= 1-C_+ e^{-m_+ r} + \cdots,\\
  h_-(r) &= C_- e^{-m_- r} + \cdots .
\end{align}
Near the origin, expanding the reduced equations around \(R=0\), one finds
\begin{align}
  K(R) &= K_2R^2 + O(R^4),\\
  a(R) &= a_MR^M + O(R^{M+2}),\\
  b(R) &= b_0 + O(R^2).
\end{align}
Here, \(C_G\), \(C_{\pm}\), \(K_2\), \(a_M\) and \(b_0\) are constants.
This near-core behavior shows that the configuration is smooth in Cartesian coordinates.

There is a useful special case at \(\eta=1\).
In this case the \(b\)-equation decouples from \(K\) and \(a\).
Moreover, the only regular finite-tension solution satisfying the boundary conditions is
\begin{align}
  b(R) \equiv 1.
\end{align}
We prove this uniqueness statement in Appendix~\ref{app:b_uniqueness_eta1}.
The remaining equations for \(K\) and \(a\) then take exactly the same form as those of the order-two family, after the identification
\begin{align}
  M \longleftrightarrow  N, \qquad a \longleftrightarrow  h .
\end{align}

\begin{figure}[t]
  \centering
  \includegraphics[width=\columnwidth]{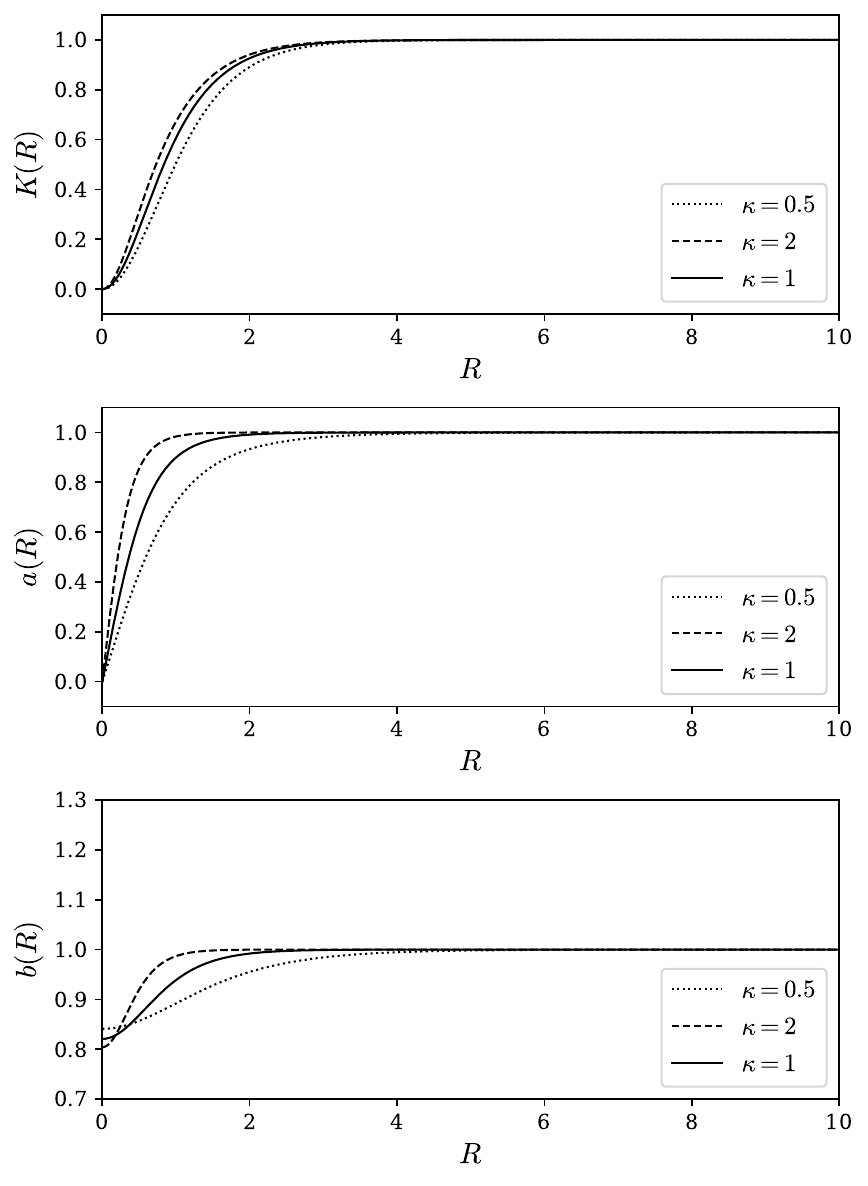}
  \caption{
    Representative numerical profiles for the order-three vortex ansatz with \(M=1\) and \(\eta=0.5\).
  The profiles are shown for \(\kappa=0.5,1,2\).
  \label{fig:order3_profiles1}
  }
\end{figure}

\begin{figure}[t]
  \centering
  \includegraphics[width=\columnwidth]{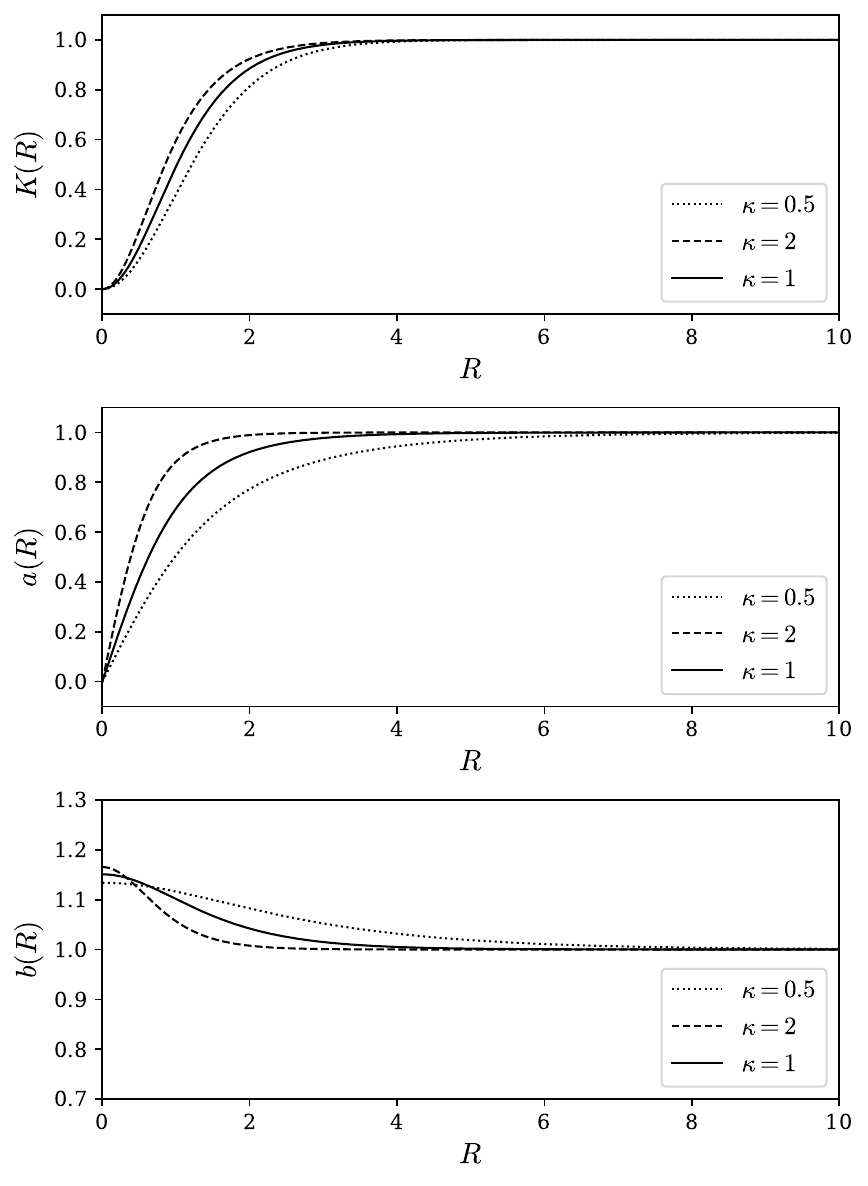}
  \caption{
    Representative numerical profiles for the order-three vortex ansatz with \(M=1\) and \(\eta=2\).
    The profiles are shown for \(\kappa=0.5,1,2\).
  \label{fig:order3_profiles2}
  }
\end{figure}

Representative numerical profiles obtained by solving the boundary value problem are shown in Figs.~\ref{fig:order3_profiles1} and \ref{fig:order3_profiles2}.

\section{Type-I, -II, and BPS-like vortices}
\label{sec:type_I_II_BPS}
We now discuss the energetics of the order-two and order-three vortex families constructed above.
Both families exhibit type-I/type-II-like energetics, with BPS-like points appearing at special values of the mass ratios.
Here we use the terms type-I, type-II, and BPS-like in analogy with the Abelian-Higgs model
\cite{Abrikosov:1956sx,Nielsen:1973cs}, or equivalently with the Ginzburg--Landau theory of superconductivity.
More precisely, type-I-like behavior means that vortices at infinite separation have higher total tension than the corresponding combined vortex, whereas type-II-like behavior means that the infinitely separated configuration has lower tension.
By a BPS-like point, we mean a point at which the reduced energy has a Bogomol'nyi-type structure and the binding energy between vortices vanishes.
As will be shown below, the BPS-like points occur at \(\kappa=1\) for the order-two family and at \((\eta,\kappa)=(1,1)\) for the order-three family.

The energy functional associated with the Lagrangian \eqref{eq:lagrangian} is
\begin{align}
  E = \int d^3x\, \Bigg[
  &\frac{1}{2e^2} \left( (\mathcal E_i^a)^2 + (\mathcal B_i^a)^2 \right) \notag \\
  &+ \frac{1}{2} \left((D_0\phi_A)^2 + (D_i\phi_A)^2\right) + V(\phi)
  \Bigg], \label{eq:energy_general}
\end{align}
where
\begin{align}
  \mathcal E_i^a := G_{0i}^a,
  \qquad
  \mathcal B_i^a := -\frac{1}{2}\varepsilon_{ijk}G_{jk}^a .
\end{align}

For the order-two family, substituting ansatz \eqref{eq:ansatz1_gauge} and \eqref{eq:ansatz1_higgs} into \eqref{eq:energy_general}, one obtains
\begin{align}
  E = \frac{\pi\mu^2}{2\lambda} \int dz\, E_2(N;\kappa),
\end{align}
where \(E_2(N;\kappa)\) is the dimensionless tension,
\begin{align}
  E_2(N;\kappa) &:= \int_0^\infty \Big[\frac{N^2 (K')^2}{4R^2} + (h')^2 \notag\\
  &+ \frac{N^2(K-1)^2h^2}{R^2} + \kappa^2 (h^2-1)^2\Big]R\,dR . \label{eq:E2_definition}
\end{align}
This functional has the same form as the energy functional of the Abelian-Higgs vortex.
The mass ratio
$\kappa=\frac{m_+}{m_G}$
in Eq.~(\ref{eq:mass-ratio})
plays the role of the Ginzburg--Landau parameter.
The integer \(N\) plays the same role as the usual Abelian-Higgs winding number.
However, in the full \(SO(3)\to A_4\) theory, \(N\) is not itself a \(\mathbb Z\)-valued topological charge.
The actual topological sector is discussed in section \ref{sec:topological_classification}.

\begin{figure}[t]
  \centering
  \includegraphics[width=\columnwidth]{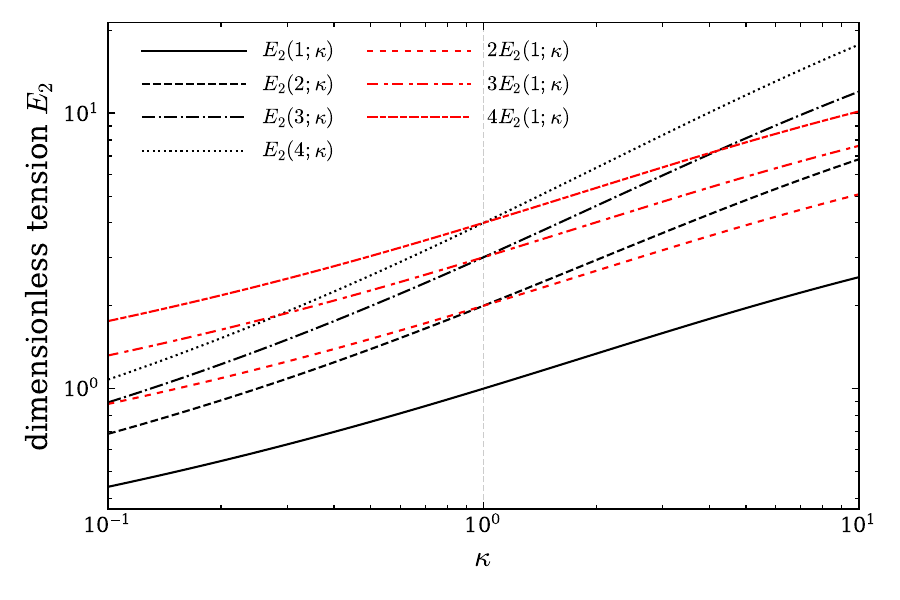}
  \caption{
  Tension comparison for the order-two family as a function of \(\kappa\).
  We compare \(E_2(N;\kappa)\), the tension of the configuration labeled by \(N\), with \(NE_2(1;\kappa)\), the total tension of \(N\) copies of the \(N=1\) configuration at infinite separation.
  \label{fig:E2_type_I_II}
  }
\end{figure}

We determine the type-I/type-II-like behavior by comparing the tension \(E_2(N;\kappa)\) of the configuration labeled by \(N\) with \(NE_2(1;\kappa)\), the tension of \(N\) copies of the \(N=1\) configuration at infinite separation.
This comparison is shown in Fig.~\ref{fig:E2_type_I_II}.
When
\begin{align}
  E_2(N;\kappa) < NE_2(1;\kappa),
\end{align}
the combined configuration has lower tension, corresponding to type-I-like behavior.
Conversely, when
\begin{align}
  E_2(N;\kappa) > NE_2(1;\kappa),
\end{align}
the infinitely separated configuration is energetically favored, corresponding to type-II-like behavior.
As seen in Fig.~\ref{fig:E2_type_I_II}, the transition between the type-I-like and type-II-like regimes occurs at \(\kappa=1\).
At this point, the reduced energy \eqref{eq:E2_definition} admits the standard Abelian-Higgs Bogomol'nyi completion, giving
\begin{align}
  E_2(N;1) = N. \label{eq:E2_BPS}
\end{align}
Thus, the binding energy between vortices vanishes at \(\kappa=1\), exactly as at the Bogomol'nyi point of the Abelian-Higgs model.

For the order-three family, substituting ansatz \eqref{eq:ansatz2_gauge} and \eqref{eq:ansatz2_higgs} into \eqref{eq:energy_general}, one obtains
\begin{align}
  E = \frac{\pi\mu^2}{2\lambda} \int dz\, E_3(M;\eta,\kappa),
\end{align}
where \(E_3(M;\eta,\kappa)\) is the dimensionless tension,
\begin{align}
  E_3(M;\eta,\kappa) &:= \frac{4}{9} \int_0^\infty \Bigg[
  \frac{M^2(K')^2}{4R^2} + (a')^2 + \frac{5}{4}(b')^2
  \nonumber\\
  &
  + \frac{M^2(K-1)^2a^2}{R^2}
  + \frac{\kappa^2(-9+4a^2+5b^2)^2}{36}
  \nonumber\\
  &
  + \frac{5\kappa^2(a^2-b^2)^2}{9\eta^2}\Bigg]R\,dR .\label{eq:E3_definition}
\end{align}
Unlike the order-two family, this functional contains two scalar profiles, \(a(R)\) and \(b(R)\).
Therefore, for generic \(\eta\), it is not equivalent to a single-component Abelian-Higgs vortex.
There is, however, a special simplification at \(\eta=1\).
In this case, regularity at the vortex core and finite tension imply
\begin{align}
  b(R)\equiv 1 ,
\end{align}
as shown in the previous section.
The dimensionless tension \eqref{eq:E3_definition} then reduces to
\begin{align}
  E_3(M;1,\kappa) &= \frac{4}{9} \int_0^\infty \Bigg[\frac{M^2(K')^2}{4R^2} + (a')^2 \notag\\
  &+ \frac{M^2(K-1)^2a^2}{R^2} + \kappa^2(a^2-1)^2\Bigg]R\,dR .\label{eq:E3_eta1}
\end{align}
This is identical to \eqref{eq:E2_definition}, up to the overall factor \(4/9\), under the identification
\begin{align}
  M \longleftrightarrow N, \qquad a \longleftrightarrow h .
\end{align}
Hence
\begin{align}
  E_3(M;1,\kappa) = \frac{4}{9}E_2(M;\kappa). \label{eq:E3_E2_relation}
\end{align}
Combining this relation with \eqref{eq:E2_BPS}, we obtain, at \(\kappa=1\),
\begin{align}
  E_3(M;1,1)=\frac{4}{9}M .
  \label{eq:E3_BPS}
\end{align}
Thus, within the order-three family, the point \((\eta,\kappa)=(1,1)\) is BPS-like:
the reduced energy functional has an Abelian-Higgs Bogomol'nyi structure, the tension is linear in the vortex label \(M\), and the binding energy between separated vortices vanishes.

\section{Topological classification}
\label{sec:topological_classification}
We now classify the vortex solutions constructed above.
A string configuration defines, at spatial infinity, a loop in the vacuum manifold \(\mathcal M=SO(3)/A_4\).
Let \(p:SU(2)\to SO(3)\) denote the double covering map. Since \(SU(2)\) is simply connected, we have
\begin{align}
  \pi_1(\mathcal M) = \pi_1(SO(3)/A_4) \simeq p^{-1}(A_4) =: \widetilde A_4 ,
\end{align}
where \(\widetilde A_4\) is the binary tetrahedral group, \(\widetilde A_4\simeq SL(2,3).\)
(See, for instance, refs.~\cite{Ishimori:2010au,Mawson:2018klj,Masaki:2023rtn}.)
Thus the loop at spatial infinity determines an element of \(\widetilde A_4\).
However, a gauge transformation at spatial infinity conjugates this element.
Therefore physical string configurations are labeled by conjugacy classes of \(\widetilde A_4\), rather than by individual elements.

Since the gauge group is \(SO(3)\), genuine Wilson lines are labeled only by integer-spin representations.
Therefore they can detect only the conjugacy class of the projected \(SO(3)\) holonomy in \(A_4\), not the full topological class in \(\widetilde A_4\).
In particular, the two classes in \(\widetilde A_4\) that project to the same class in \(A_4\) cannot be distinguished by such Wilson loops,
although they may represent different topological sectors of the Higgs configuration.

We now make explicit the relation between the projected \(SO(3)\) holonomy and the topological class in
\(\widetilde A_4=p^{-1}(A_4)\simeq \pi_1(SO(3)/A_4)\).

Throughout this section, \([g]\) denotes the conjugacy class of \(g\).
With our convention, the conjugacy classes of \(A_4\) are
\begin{equation}
  \begin{aligned}
  C_0 &= [1],        & |C_0|=1,\\
  C_1 &= [r_2],      & |C_1|=3,\\
  C_2 &= [r_3^2],    & |C_2|=4,\\
  C_3 &= [r_3],      & |C_3|=4,
  \end{aligned}
\end{equation}
where \(r_2\) and \(r_3\) are defined in Eq.~\eqref{eq:r2r3}, and
\(|C|\) denotes the number of elements in a conjugacy class \(C\).
The class \(C_1\) consists of the order-two elements, while \(C_2\) and
\(C_3\) are the two conjugacy classes of order-three elements.

Let \(\widetilde r_2,\widetilde r_3\in\widetilde A_4\) be lifts of \(r_2,r_3\), respectively:
\begin{align}
  p(\widetilde r_2)=r_2,
  \qquad
  p(\widetilde r_3)=r_3.
\end{align}
We choose them so that
\begin{align}
  \widetilde r_2^2=-1,
  \qquad
  \widetilde r_3^3=-1.
\end{align}
Thus \(\widetilde r_2\) has order four, and \(\widetilde r_3\) has order six.
The conjugacy classes of \(\widetilde A_4\) are
\begin{equation}
  \begin{aligned}
  D_0 &= [1],                  &  |D_0|=1, \\
  D_1 &= [-1],                 &  |D_1|=1, \\
  D_2 &= [\widetilde r_3],     &  |D_2|=4, \\
  D_3 &= [\widetilde r_3^{-1}],&  |D_3|=4, \\
  D_4 &= [\widetilde r_3^2],   &  |D_4|=4, \\
  D_5 &= [\widetilde r_3^{-2}],&  |D_5|=4, \\
  D_6 &= [\widetilde r_2],     &  |D_6|=6.
  \end{aligned}
\end{equation}
Under the covering map \(p:\widetilde A_4\to A_4\), these classes project as
\begin{equation}
  \begin{split}
  D_0,D_1 &\longmapsto C_0,\\
  D_6 &\longmapsto C_1,\\
  D_3,D_4 &\longmapsto C_2,\\
  D_2,D_5 &\longmapsto C_3.
  \end{split}
\end{equation}

For the order-two and order-three vortex families constructed above, the asymptotic \(SO(3)\) holonomies are
\begin{align}
  U_N=r_2^N,
  \qquad
  U_M=r_3^M,
\end{align}
respectively.
Their topological classes are represented by \(\widetilde r_2^{\,N}\) and \(\widetilde r_3^{\,M}\).
Since \(\widetilde r_2\) and \(\widetilde r_3\) have orders four and six, respectively, the classification depends on \(N\mod 4\) and \(M\mod 6\).
The result is summarized in TABLE~\ref{tab:topological_classification}.

In particular, the projected \(SO(3)\) holonomy is trivial for \(N=2\pmod 4\) and \(M=3\pmod 6\), but the corresponding topological class is \(D_1=[-1]\).
These configurations are therefore topologically nontrivial despite having trivial \(SO(3)\) holonomy.

\begin{table}[t]
\centering
\begin{tabular}{c|c|c|c}
       &              & \(SO(3)\)     & vortex\\
family & vortex label &holonomy class & topological sector \\
\hline
order-two & \(N=0\!\!\pmod 4\) & \(C_0\) & \(D_0\) \\
order-two & \(N=1\!\!\pmod 4\) & \(C_1\) & \(D_6\) \\
order-two & \(N=2\!\!\pmod 4\) & \(C_0\) & \(D_1\) \\
order-two & \(N=3\!\!\pmod 4\) & \(C_1\) & \(D_6\) \\
\hline
order-three & \(M=0\!\!\pmod 6\) & \(C_0\) & \(D_0\) \\
order-three & \(M=1\!\!\pmod 6\) & \(C_3\) & \(D_2\) \\
order-three & \(M=2\!\!\pmod 6\) & \(C_2\) & \(D_4\) \\
order-three & \(M=3\!\!\pmod 6\) & \(C_0\) & \(D_1\) \\
order-three & \(M=4\!\!\pmod 6\) & \(C_3\) & \(D_5\) \\
order-three & \(M=5\!\!\pmod 6\) & \(C_2\) & \(D_3\)
\end{tabular}
\caption{
Topological classification of the two vortex families.
The \(SO(3)\) holonomy is classified by conjugacy classes \(C_i\subset A_4\), while the full topological sector is classified by conjugacy classes \(D_i\subset \widetilde A_4=p^{-1}(A_4)\).
\label{tab:topological_classification}
}
\end{table}

\section{Relation to non-invertible one-form symmetries in \(A_4\) gauge theory}
\label{sec:noninvertible_symmetry}
We now explain the relation between the vortices constructed above and the non-invertible one-form symmetry of the infrared \(A_4\) gauge theory.
At energies well below the mass scales \(m_+\), \(m_-\), and \(m_G\), the massive Higgs and gauge fluctuations are integrated out, and the \(SO(3)\to A_4\) Higgs phase is described by a discrete \(A_4\) gauge theory.
It is known that a finite-group gauge theory has a non-invertible one-form symmetry, whose topological surface operators are labeled by conjugacy classes of the gauge group~\cite{Gukov:2008sn,Heidenreich:2021xpr}.
These surface operators are often called Gukov--Witten operators.

Let us describe these operators in lattice language.
In a lattice formulation of an \(A_4\) gauge theory, a group element
\begin{align}
  U_{yx} \in A_4
\end{align}
is assigned to each oriented link \((y,x)\) from \(x\) to \(y\) with
\begin{align}
  U_{xy} = U_{yx}^{-1}.
\end{align}
For an oriented plaquette
\begin{align}
  \square = (x_0x_3)(x_3x_2)(x_2x_1)(x_1x_0),
\end{align}
one imposes the flatness condition
\begin{align}
  \operatorname{Hol}_{x_0}(\square) := U_{x_0x_3}U_{x_3x_2}U_{x_2x_1}U_{x_1x_0} = 1_{A_4},
\end{align}
where \(1_{A_4}\) is the identity element of \(A_4\).
The quantity \(\operatorname{Hol}_{x_0}(\square)\) is the finite-group analogue of the continuum holonomy
\begin{align}
  \mathrm{P}\exp\left(-i\oint_{\partial \square} W_\mu dx^\mu\right).
\end{align}

Now let \(\Sigma\) be a codimension-two surface in spacetime, and let \(\square\) be a plaquette linking \(\Sigma\).
For a fixed element \(g\in A_4\), we define \(T_g(\Sigma;x_0)\) as the insertion which changes the flatness condition on the linking plaquette to
\begin{align}
  \operatorname{Hol}_{x_0}(\square) = g .
\end{align}
Equivalently, \(T_g(\Sigma;x_0)\) inserts a prescribed \(A_4\) flux along \(\Sigma\).

The operator \(T_g(\Sigma;x_0)\) is not a physical surface operator by itself, because the holonomy is conjugated under a gauge transformation or a change of base point.
Indeed, under a lattice gauge transformation
\begin{align}
  U_{xy} \longmapsto s_xU_{xy}s_y^{-1}, \qquad s_x \in A_4,
\end{align}
the plaquette holonomy transforms as
\begin{align}
  \operatorname{Hol}_{x_0}(\square) \longmapsto s_{x_0}\operatorname{Hol}_{x_0}(\square)s_{x_0}^{-1}.
\end{align}
Similarly, if we change the base point, for example, from \(x_0\) to \(x_1\),
then the same plaquette holonomy is written as
\begin{align}
  \operatorname{Hol}_{x_1}(\square)
  &= U_{x_1x_0}U_{x_0x_3}U_{x_3x_2}U_{x_2x_1} \notag\\
  &= U_{x_0x_1}^{-1} \operatorname{Hol}_{x_0}(\square) U_{x_0x_1}.
\end{align}
Thus, changing the base point conjugates the plaquette holonomy.

As shown above, both gauge transformations and changes of base point conjugate the plaquette holonomy.
Therefore, for a conjugacy class \(C=[g]\subset A_4\), we can define a gauge-invariant and base-point-independent operator by summing over \(C\):
\begin{align}
  T_C(\Sigma) := \sum_{h\in C}T_h(\Sigma;x_0).
\end{align}
Indeed, conjugation merely permutes the elements of \(C\), and hence only relabels the terms on the right-hand side, leaving the sum unchanged.
The operators \(T_C(\Sigma)\) are the Gukov--Witten operators.
They
are topological
in the sense that the holonomy generated by \(T_C(\Sigma)\)
is preserved under continuous deformations of \(\Sigma\)
which do not change the linking number between \(\Sigma\) and the plaquette \(\square\).
Since the holonomy operator is
a one-dimensional object, the
operators
 \(T_C(\Sigma)\) are symmetry generators of the one-form symmetry~\cite{Heidenreich:2021xpr}.
 Note that this one-form symmetry can be explicitly broken by adding dynamical matter fields which can screen
 the holonomy.

The fusion rules of these operators are given by the conjugacy class algebra of \(A_4\).
Using the notation for conjugacy classes introduced in section \ref{sec:topological_classification}, the result is summarized in TABLE~\ref{tab:A4_class_algebra}.
Since fusing two such operators does not generally produce a single operator but rather a sum of operators, the corresponding one-form symmetry is non-invertible.

\begin{table}[h]
  \centering
  \begin{tabular}{c|cccc}
      \(\times\)  & \(T_{C_0}\) & \(T_{C_1}\) & \(T_{C_2}\) & \(T_{C_3}\) \\ \hline
      \(T_{C_0}\) & \(T_{C_0}\) &\(T_{C_1}\) &\(T_{C_2}\) &\(T_{C_3}\) \\
      \(T_{C_1}\) & \(T_{C_1}\) &\(3 T_{C_0} + 2 T_{C_1}\) &\(3 T_{C_2}\) &\(3 T_{C_3}\) \\
      \(T_{C_2}\) & \(T_{C_2}\) &\(3 T_{C_2}\) &\(4 T_{C_3}\) &\(4 T_{C_0} + 4 T_{C_1}\) \\
      \(T_{C_3}\) & \(T_{C_3}\) &\(3 T_{C_3}\) &\(4 T_{C_0} + 4 T_{C_1}\) &\(4 T_{C_2}\) \\
    \end{tabular}
  \caption{
    Fusion rules of Gukov--Witten surface operators in \(A_4\) gauge theory.
    Here \(C_0,C_1,C_2,C_3\) are the conjugacy classes of \(A_4\) defined in section \ref{sec:topological_classification}.
    A similar fusion rule can be found
    in refs.~\cite{Mawson:2018klj,Masaki:2023rtn}
    for $G=U(1) \times SO(3)$ with spin-2 complex Higgs fields.
    \label{tab:A4_class_algebra}
  }
\end{table}

We now compare these infrared surface operators with the smooth vortices in the \(SO(3)\to A_4\) Higgs theory.
As explained in section \ref{sec:topological_classification}, the smooth finite-tension vortices are classified by conjugacy classes of \(\widetilde A_4 = \pi_1(SO(3)/A_4)\).
This classification contains UV information about how the fields wind in the continuous vacuum manifold \(SO(3)/A_4\).
In the infrared \(A_4\) gauge theory, however, this detailed winding information is not retained in the effective description:
only the \(A_4\) holonomy around a linking circle remains, while the information distinguishing different lifts to \(\widetilde A_4\) is lost.

This observation naturally leads to a map from the conjugacy classes of \(\widetilde A_4\) to those of \(A_4\).
Using the notation of section \ref{sec:topological_classification}, the covering map
\(p:SU(2)\to SO(3)\), restricted to \(\widetilde A_4\subset SU(2)\), induces a map on conjugacy classes,
\begin{align}
  [\widetilde g]\longmapsto [p(\widetilde g)] .
\end{align}
A smooth vortex in the sector \([\widetilde g]\) is therefore described in the infrared by the Gukov--Witten operator \(T_{[p(\widetilde g)]}\).
In other words, different \(\widetilde A_4\) sectors may become indistinguishable after flowing to the infrared \(A_4\) gauge theory, if they have the same projected \(A_4\) holonomy.
For example, the two sectors \(D_3\) and \(D_4\) both project to \(C_2\).
Hence, vortices in either sector are described in the infrared by the same Gukov--Witten operator \(T_{C_2}\).
These two sectors are distinguished by the smooth-core structure of the \(SO(3)\to A_4\) Higgs configuration, but this distinction is not visible in the infrared \(A_4\) gauge theory.

Thus, the magnetic flux carried by each smooth finite-tension vortex constructed in this paper flows in the infrared to a Gukov--Witten surface operator in the \(A_4\) gauge theory.
In this sense, the \(SO(3)\) gauge--Higgs model provides a smooth finite-tension UV realization of these infrared surface operators within a renormalizable field theory.

\section{Discussion}
\label{sec:discussion}
In this article, the order-two and order-three vortex families have been analyzed within minimal cylindrically symmetric ans\"atze.
A notable feature of these solutions is the appearance of BPS-like points:
\(\kappa=1\) for the order-two family and \((\eta,\kappa)=(1,1)\) for the order-three family.
At these points, the reduced equations and energy functionals take the same form as those of the Abelian-Higgs model at its Bogomol'nyi point.
It follows that static multi-vortex configurations can exist with no binding energy, as in the ordinary BPS Abelian-Higgs theory.
In this sense, the vortices constructed above are BPS-like.
This BPS-like property, however, is a statement within the reduced ansatz:
although the energy saturates the corresponding Bogomol'nyi bound in this sector, fluctuations outside the ansatz may still lower the energy.
This question is directly related to the stability of the BPS-like vortices constructed here, and we leave it for future work.

We next clarify how the global structure of the gauge group affects the interpretation of the vortices.
At the local level, one may replace \(SO(3)\) by \(SU(2)\), since the two theories have the same Lie algebra.
The local Lagrangian and the radial vortex equations are therefore unchanged.
The global interpretation, however, is different.
In the \(SU(2)\) theory, the Higgs phase is
\begin{align}
  SU(2) \longrightarrow \widetilde A_4 ,
\end{align}
rather than \(SO(3)\to A_4\).
Consequently, the infrared discrete gauge theory is a \(\widetilde A_4\) gauge theory.

It is also natural to ask how this construction extends to other finite subgroups of \(SO(3)\).
The same strategy should apply to other Higgs phases
\begin{align}
    SO(3)\to\Gamma,
\end{align}
where \(\Gamma\) is a finite subgroup of \(SO(3)\).
In such phases, one expects smooth vortices whose asymptotic holonomies lie in \(\Gamma\), in direct analogy with the \(A_4\) vortices constructed in this article.
For example, even within the spin-3 representation, a different choice of renormalizable \(SO(3)\)-invariant potential can lead to the dihedral group \(D_3\simeq S_3\) as the residual symmetry.
Other polyhedral groups can be realized by using higher-spin Higgs fields.
In particular, spin-4 and spin-6 Higgs fields, together with suitable renormalizable \(SO(3)\)-invariant potentials, naturally realize \(S_4\) and \(A_5\) as residual symmetries, respectively~\cite{Ovrut:1977cn,Etesi:1996urw,King:2018fke}.
For a general finite subgroup \(\Gamma\subset SO(3)\), the topological classification of smooth vortices takes the universal form
\begin{align}
    \pi_1(SO(3)/\Gamma) = p^{-1}(\Gamma)\subset SU(2),
\end{align}
where \(p:SU(2)\to SO(3)\) is the double covering map.
Thus, the UV topological sectors are classified by the binary lift of the residual finite group.
It would be interesting to understand whether the BPS-like behavior found in this article persists for other choices of \(\Gamma\).
The existence of such a point is not automatic.
Unlike the topological classification above, it depends on the detailed structure of the Higgs representation and the choice of invariant potential.

There are several avenues for future work.
First, an application to cosmic strings
is one of the important future directions.
When two cosmic strings
characterized by commutative elements of
the fundamental group
collide, they pass through or reconnect with each other.
Thus, cosmic strings can reduce their number, reaching the standard scaling law.
On the other hand,
when two cosmic strings corresponding to non-commutative elements of
the fundamental group collide,
a bridge is created between them
\cite{Poenaru1977,Mermin:1979zz,Kobayashi:2008pk,Kobayashi:2024aip}.
As a consequence,
the number of cosmic strings is not reduced during cosmic expansion, and
a cosmic string network is inevitably formed \cite{McGraw:1997nx},
in which $S_3$ cosmic strings were explicitly discussed.
Our construction provides a concrete framework for studying the dynamics, fusion, and interactions of non-Abelian cosmic strings in a fully renormalizable gauge theory.

$A_4$ vortices can be obtained in a $U(1) \times SO(3) $ gauge theory with spin-2 (complex) Higgs fields \cite{Semenoff:2006vv,Kobayashi:2008pk,Mawson:2018klj,Kobayashi:2024aip} instead of spin-3 Higgs fields studied in this article.
In such a case,
the fundamental group is
some product of $A_4$ and ${\mathbb Z}$ \cite{Kobayashi:2011xb}.
Consequently,
BPS $A_4$ vortices would satisfy
the mass-charge relations,
and would be
1/2 BPS states
when embedded into supersymmetric gauge theory.
An analogous case is
that a $U(1) \times SO(3)$ supersymmetric gauge theory
with
spin-1 (complex) Higgs fields
admits 1/2 BPS Alice strings
\cite{Chatterjee:2017jsi,Chatterjee:2017hya},\footnote{
See refs.~\cite{Leonhardt:2000km,Sato:2018nqy,Chatterjee:2019rch} for  global analogues for these strings.
}
as an extension of an $SO(3)$ gauge theory with
spin-2 (real) Higgs fields
\cite{Schwarz:1982ec}.

When non-invertible defects create linkings,
a non-invertible symmetry gives a selection rule
for linkings,
as explicitly shown for axion electrodynamics
\cite{Hidaka:2024kfx}.
As compact non-invertible defects,
non-Abelian vortex knots and links have thus far been constructed ad hoc
\cite{Brekke:1992he,Annala:2022bdd,Rajamaki:2023ymv,Kobayashi:2024aip}.
In particular, in ref.~\cite{Kobayashi:2024aip},
all possible knots and links are classified for $A_4$ vortices.
The non-invertible symmetry may provide a general framework for the construction of such compact non-invertible defects.

Finally, there are many important directions,
such as  quantum computation,
finite temperature, finite density,
monopole attachment, and
non-invertible symmetry.

%%%%%%%%%%%%%%%%%%%%%%%
\section*{Acknowledgments}
\noindent
This work is supported by JST Grant No. JPMJPF2221 (YA), by JSPS KAKENHI Grant No. JP22K03601 (TH),
JP22H01221 and JP23K22492 (MN),
JP25K17394 (RY), and by
the WPI program ``Sustainability with Knotted
 Chiral Meta Matter (WPI-SKCM$^2$)'' at Hiroshima University (MN).

\appendix
\section{Lie algebra generators in the spin-3 representation}
\label{app:spin3_generators}
In the orthonormal basis \(e_1,\ldots,e_7\) of \(\mathcal H_3\),
the representation matrices \(X_a\) of the Lie algebra in the spin-3 representation are given by
\begin{align}
    X_1 &= \begin{pmatrix}
        0&0&0&0&0&0&0\\
        0&0&-3/2&0&0&-\sqrt{15}/2&0\\
        0&3/2&0&0&-\sqrt{15}/2&0&0\\
        0&0&0&0&0&0&2\\
        0&0&\sqrt{15}/2&0&0&1/2&0\\
        0&\sqrt{15}/2&0&0&-1/2&0&0\\
        0&0&0&-2&0&0&0
    \end{pmatrix},\\
    X_2 &= \begin{pmatrix}
        0&0&3/2&0&0&-\sqrt{15}/2&0\\
        0&0&0&0&0&0&0\\
        -3/2&0&0&-\sqrt{15}/2&0&0&0\\
        0&0&\sqrt{15}/2&0&0&-1/2&0\\
        0&0&0&0&0&0&2\\
        \sqrt{15}/2&0&0&1/2&0&0&0\\
        0&0&0&0&-2&0&0
    \end{pmatrix},\\
    X_3 &= \begin{pmatrix}
        0&-3/2&0&0&-\sqrt{15}/2&0&0\\
        3/2&0&0&-\sqrt{15}/2&0&0&0\\
        0&0&0&0&0&0&0\\
        0&\sqrt{15}/2&0&0&1/2&0&0\\
        \sqrt{15}/2&0&0&-1/2&0&0&0\\
        0&0&0&0&0&0&2\\
        0&0&0&0&0&-2&0
    \end{pmatrix}.
    \\\notag
\end{align}

\section{Explicit form of the quartic invariant}
\label{app:quartic_invariant}
In this appendix we give the explicit component expression of the quartic \(SO(3)\)-invariant polynomial \(Q\) defined in section \ref{sec:model}.
Writing
\begin{align}
  \phi=\sum_{A=1}^{7}\phi_A e_A ,
\end{align}
one finds
\begin{widetext}
    \begin{align}
        Q(\phi) = \frac{1}{675}\Biggl(
            &8 \left(- 2 \phi_{1}^{2} + \phi_{2}^{2} - \sqrt{15} \phi_{2} \phi_{5} + \phi_{3}^{2} + \sqrt{15} \phi_{3} \phi_{6}\right)^{2}
            + 8 \left(\phi_{1}^{2} - \sqrt{15} \phi_{1} \phi_{4} + \phi_{2}^{2} + \sqrt{15} \phi_{2} \phi_{5} - 2 \phi_{3}^{2}\right)^{2}
            \nn\\
            +&8 \left(\phi_{1}^{2} + \sqrt{15} \phi_{1} \phi_{4} - 2 \phi_{2}^{2} + \phi_{3}^{2} - \sqrt{15} \phi_{3} \phi_{6}\right)^{2}
            +\left(3 \phi_{1} \phi_{2} + \sqrt{15} \phi_{1} \phi_{5} - \sqrt{15} \phi_{2} \phi_{4} + 4 \sqrt{15} \phi_{3} \phi_{7} + 15 \phi_{4} \phi_{5}\right)^{2}
            \nn\\
            +&\left(3 \phi_{1} \phi_{3} - \sqrt{15} \phi_{1} \phi_{6} + 4 \sqrt{15} \phi_{2} \phi_{7} + \sqrt{15} \phi_{3} \phi_{4} + 15 \phi_{4} \phi_{6}\right)^{2}
            \nn\\
            +&\left(4 \sqrt{15} \phi_{1} \phi_{7} + 3 \phi_{2} \phi_{3} + \sqrt{15} \phi_{2} \phi_{6} - \sqrt{15} \phi_{3} \phi_{5} + 15 \phi_{5} \phi_{6}\right)^{2}
        \Biggr).
    \end{align}
\end{widetext}
Since the right-hand side is written as a sum of squares, this expression makes the non-negativity of \(Q\) manifest:
\begin{align}
  Q(\phi)\ge 0 .
\end{align}
It is also useful to relate this expression to the standard tensorial description of the spin-3 representation.
An element of \(\mathcal H_3\) can equivalently be represented by a totally symmetric traceless rank-three tensor \(D_{ijk}\).
Here ``traceless'' means that the contraction of any two indices with the Euclidean metric vanishes:
\begin{align}
    \delta^{ij}D_{ijk}=0 .
\end{align}
Because \(D_{ijk}\) is totally symmetric, this condition is independent of which pair of indices is contracted.
The corresponding harmonic polynomial is
\begin{align}
  \phi(\mathcal X) = D_{ijk} \mathcal X_i \mathcal X_j \mathcal X_k,
  \qquad
  \mathcal X=(\mathcal X_1,\mathcal X_2,\mathcal X_3)=(\mathcal X,\mathcal Y,\mathcal Z),
\end{align}
where repeated indices are summed over \(i,j,k=1,2,3\).
The traceless condition is precisely the condition that this polynomial is harmonic, \(\Delta\phi=0\).
Explicitly, in the basis used in this paper,
\begin{equation}
  \begin{split}
  D_{111} &= \sqrt{7} \phi_{1}, \\
  D_{112} &= - \frac{\sqrt{7}}{2}\phi_{2} - \frac{\sqrt{105}}{6}\phi_{5}, \\
  D_{113} &= - \frac{\sqrt{7}}{2}\phi_{3} + \frac{\sqrt{105}}{6}\phi_{6}, \\
  D_{122} &= - \frac{\sqrt{7}}{2}\phi_{1} + \frac{\sqrt{105}}{6}\phi_{4}, \\
  D_{123} &= \frac{\sqrt{105}}{6}\phi_{7}, \\
  D_{133} &= - \frac{\sqrt{7}}{2}\phi_{1} - \frac{\sqrt{105}}{6}\phi_{4}, \\
  D_{222} &= \sqrt{7} \phi_{2}, \\
  D_{223} &= - \frac{\sqrt{7}}{2}\phi_{3} - \frac{\sqrt{105}}{6}\phi_{6}, \\
  D_{233} &= - \frac{\sqrt{7}}{2}\phi_{2} + \frac{\sqrt{105}}{6}\phi_{5}, \\
  D_{333} &= \sqrt{7} \phi_{3},
  \end{split}
\end{equation}
with all other components determined by total symmetry,
\(D_{ijk}=D_{(ijk)}\).

Under the above identification between \(\phi\) and \(D_{ijk}\), the invariants used in section \ref{sec:model} are related to the tensor invariants by
\begin{align}
  I_2(\phi) &= \frac{2}{35}J_2(D),\\
  Q(\phi) &= \frac{8}{11025} \left(3J_4(D)-J_2(D)^2\right),
\end{align}
where
\begin{align}
  J_2(D) &:= D_{ijk}D_{ijk},\\
  J_4(D) &:= D_{ijk}D_{ij\ell}D_{pqk}D_{pq\ell}.
\end{align}

\section{Uniqueness of the \(b\)-profile at \(\eta=1\)}
\label{app:b_uniqueness_eta1}
In this appendix we prove that, at \(\eta=1\), the \(b\)-profile in the order-three vortex ansatz is uniquely fixed by the boundary conditions required for finite tension and regularity.

At \(\eta=1\), the equation of motion for \(b\) decouples from the other profile functions and becomes
\begin{align}
  b'' = -\frac{b'}{R}+2\kappa^2(b^2-1)b. \label{eq:b_equation_eta1}
\end{align}
The boundary conditions for \(b\) are
\begin{align}
  b'(0)=0,\qquad b(\infty)=1 .
\end{align}
The constant profile \(b(R)\equiv1\) is clearly a solution.
We now show that it is the only finite-tension solution satisfying these boundary conditions.

Define
\begin{align}
  T(R) := \frac{1}{2}(b')^2 - \frac{\kappa^2}{2}(b^2-1)^2.
\end{align}
Using \eqref{eq:b_equation_eta1}, we find
\begin{align}
  T' &= b'b'' - 2\kappa^2b(b^2-1)b'
  = - \frac{(b')^2}{R} \le 0 .
\end{align}
Thus \(T(R)\) is non-increasing.
At the origin, the boundary condition gives
\begin{align}
  T(0) = - \frac{\kappa^2(b(0)^2-1)^2}{2} \le 0 .
\end{align}
On the other hand, finite tension implies \(b(R)\to1\) and \(b'(R)\to0\) as \(R\to\infty\), as follows from the \(b\)-dependent potential term and the positive \((b')^2\) term in \eqref{eq:E3_definition}.
Hence \(T(\infty)=0\).
Since \(T\) is non-increasing, we have
\begin{align}
  0 = T(\infty)\le T(R)\le T(0)\le0 .
\end{align}
Therefore \(T(R)\equiv0\). In particular,
\begin{align}
  T'(R) = -\frac{(b')^2}{R}=0
\end{align}
for all \(R>0\). Hence \(b'(R)\equiv0\), so \(b\) is constant. The boundary
condition \(b(\infty)=1\) then gives \(b(R)\equiv1\).

This proves the uniqueness of the \(b\)-profile at \(\eta=1\).

%% References
\newcommand{\arxivfont}{\rmfamily}
\bibliographystyle{yautphysm}
\bibliography{draft.bib}

\providecommand{\href}[2]{#2}
\providecommand{\arxivfont}{\ttfamily}
\begingroup\raggedright
\begin{thebibliography}{10}

\bibitem{Vilenkin:2000jqa}
A.~Vilenkin and E.~P.~S. Shellard, {\itshape {Cosmic Strings and Other Topological Defects}}.
\newblock Cambridge University Press, 7, 2000.

\bibitem{Manton:2004tk}
N.~S. Manton and P.~Sutcliffe, \href{http://dx.doi.org/10.1017/CBO9780511617034}{{\itshape {Topological solitons}}}.
\newblock Cambridge Monographs on Mathematical Physics. Cambridge University Press, 2004.

\bibitem{Weinberg:2012pjx}
E.~J. Weinberg, \href{http://dx.doi.org/10.1017/CBO9781139017787}{{\itshape {Classical solutions in quantum field theory}: {Solitons and Instantons in High Energy Physics}}}.
\newblock Cambridge Monographs on Mathematical Physics. Cambridge University Press, 9, 2012.

\bibitem{Abrikosov:1956sx}
A.~A. Abrikosov, {\itshape {On the Magnetic Properties of Superconductors of the Second Group}}, Sov. Phys. JETP {\bfseries 5} (1957) 1174--1182.

\bibitem{Nielsen:1973cs}
H.~B. Nielsen and P.~Olesen, {\itshape {Vortex Line Models for Dual Strings}}, \href{http://dx.doi.org/10.1016/0550-3213(73)90350-7}{Nucl. Phys. B {\bfseries 61} (1973) 45--61}.

\bibitem{Bais:1980vd}
F.~A. Bais, {\itshape {FLUX METAMORPHOSIS}}, \href{http://dx.doi.org/10.1016/0550-3213(80)90474-5}{Nucl. Phys. B {\bfseries 170} (1980) 32--43}.

\bibitem{Alford:1990mk}
M.~G. Alford, K.~Benson, S.~R. Coleman, J.~March-Russell, and F.~Wilczek, {\itshape {The Interactions and Excitations of Nonabelian Vortices}}, \href{http://dx.doi.org/10.1103/PhysRevLett.64.1632}{Phys. Rev. Lett. {\bfseries 64} (1990) 1632}. [Erratum: Phys.Rev.Lett. 65, 668 (1990)].

\bibitem{Wilczek:1989kn}
F.~Wilczek and Y.-S. Wu, {\itshape Space-time approach to holonomy scattering}, \href{http://dx.doi.org/10.1103/PhysRevLett.65.13}{Phys. Rev. Lett. {\bfseries 65} (1990) 13--16}.

\bibitem{Alford:1990ur}
M.~G. Alford, K.~Benson, S.~R. Coleman, J.~March-Russell, and F.~Wilczek, {\itshape {Zero modes of nonabelian vortices}}, \href{http://dx.doi.org/10.1016/0550-3213(91)90331-Q}{Nucl. Phys. B {\bfseries 349} (1991) 414--438}.

\bibitem{Alford:1992yx}
M.~G. Alford, K.-M. Lee, J.~March-Russell, and J.~Preskill, {\itshape {Quantum field theory of nonAbelian strings and vortices}}, \href{http://dx.doi.org/10.1016/0550-3213(92)90468-Q}{Nucl. Phys. B {\bfseries 384} (1992) 251--317} [\href{http://arxiv.org/abs/hep-th/9112038}{{\arxivfont hep-th/9112038}}].

\bibitem{Bucher:1991bc}
M.~Bucher, K.-M. Lee, and J.~Preskill, {\itshape {On detecting discrete Cheshire charge}}, \href{http://dx.doi.org/10.1016/0550-3213(92)90174-A}{Nucl. Phys. B {\bfseries 386} (1992) 27--42} [\href{http://arxiv.org/abs/hep-th/9112040}{{\arxivfont hep-th/9112040}}].

\bibitem{Lo:1993hp}
H.-K. Lo and J.~Preskill, {\itshape {NonAbelian vortices and nonAbelian statistics}}, \href{http://dx.doi.org/10.1103/PhysRevD.48.4821}{Phys. Rev. D {\bfseries 48} (1993) 4821--4834} [\href{http://arxiv.org/abs/hep-th/9306006}{{\arxivfont hep-th/9306006}}].

\bibitem{Brekke:1992he}
L.~Brekke, H.~Dykstra, S.~J. Hughes, and T.~D. Imbo, {\itshape {Knots and links of nonabelian string}}, \href{http://dx.doi.org/10.1016/0370-2693(92)91103-G}{Phys. Lett. B {\bfseries 288} (1992) 273--278}.

\bibitem{Brekke:1997jj}
L.~Brekke, S.~J. Collins, and T.~D. Imbo, {\itshape {NonAbelian vortices on surfaces and their statistics}}, \href{http://dx.doi.org/10.1016/S0550-3213(97)00409-4}{Nucl. Phys. B {\bfseries 500} (1997) 465--485} [\href{http://arxiv.org/abs/hep-th/9701056}{{\arxivfont hep-th/9701056}}].

\bibitem{McGraw:1997nx}
P.~McGraw, {\itshape {Evolution of a nonAbelian cosmic string network}}, \href{http://dx.doi.org/10.1103/PhysRevD.57.3317}{Phys. Rev. D {\bfseries 57} (1998) 3317--3339} [\href{http://arxiv.org/abs/astro-ph/9706182}{{\arxivfont astro-ph/9706182}}].

\bibitem{Hanany:2003hp}
A.~Hanany and D.~Tong, {\itshape {Vortices, instantons and branes}}, \href{http://dx.doi.org/10.1088/1126-6708/2003/07/037}{JHEP {\bfseries 07} (2003) 037} [\href{http://arxiv.org/abs/hep-th/0306150}{{\arxivfont hep-th/0306150}}].

\bibitem{Auzzi:2003fs}
R.~Auzzi, S.~Bolognesi, J.~Evslin, K.~Konishi, and A.~Yung, {\itshape {NonAbelian superconductors: Vortices and confinement in N=2 SQCD}}, \href{http://dx.doi.org/10.1016/j.nuclphysb.2003.09.029}{Nucl. Phys. B {\bfseries 673} (2003) 187--216} [\href{http://arxiv.org/abs/hep-th/0307287}{{\arxivfont hep-th/0307287}}].

\bibitem{Eto:2005yh}
M.~Eto, Y.~Isozumi, M.~Nitta, K.~Ohashi, and N.~Sakai, {\itshape {Moduli space of non-Abelian vortices}}, \href{http://dx.doi.org/10.1103/PhysRevLett.96.161601}{Phys. Rev. Lett. {\bfseries 96} (2006) 161601} [\href{http://arxiv.org/abs/hep-th/0511088}{{\arxivfont hep-th/0511088}}].

\bibitem{Eto:2006cx}
M.~Eto, K.~Konishi, G.~Marmorini, M.~Nitta, K.~Ohashi, W.~Vinci, and N.~Yokoi, {\itshape {Non-Abelian Vortices of Higher Winding Numbers}}, \href{http://dx.doi.org/10.1103/PhysRevD.74.065021}{Phys. Rev. D {\bfseries 74} (2006) 065021} [\href{http://arxiv.org/abs/hep-th/0607070}{{\arxivfont hep-th/0607070}}].

\bibitem{Eto:2006pg}
M.~Eto, Y.~Isozumi, M.~Nitta, K.~Ohashi, and N.~Sakai, {\itshape {Solitons in the Higgs phase: The Moduli matrix approach}}, \href{http://dx.doi.org/10.1088/0305-4470/39/26/R01}{J. Phys. A {\bfseries 39} (2006) R315--R392} [\href{http://arxiv.org/abs/hep-th/0602170}{{\arxivfont hep-th/0602170}}].

\bibitem{Shifman:2009zz}
M.~Shifman and A.~Yung, \href{http://dx.doi.org/10.1017/9781009402200}{{\itshape {Supersymmetric Solitons}}}.
\newblock Cambridge University Press, 2009.

\bibitem{Balachandran:2005ev}
A.~P. Balachandran, S.~Digal, and T.~Matsuura, {\itshape {Semi-superfluid strings in high density QCD}}, \href{http://dx.doi.org/10.1103/PhysRevD.73.074009}{Phys. Rev. D {\bfseries 73} (2006) 074009} [\href{http://arxiv.org/abs/hep-ph/0509276}{{\arxivfont hep-ph/0509276}}].

\bibitem{Nakano:2007dr}
E.~Nakano, M.~Nitta, and T.~Matsuura, {\itshape {Non-Abelian strings in high density QCD: Zero modes and interactions}}, \href{http://dx.doi.org/10.1103/PhysRevD.78.045002}{Phys. Rev. D {\bfseries 78} (2008) 045002} [\href{http://arxiv.org/abs/0708.4096}{{\arxivfont 0708.4096 [hep-ph]}}].

\bibitem{Eto:2009kg}
M.~Eto and M.~Nitta, {\itshape {Color Magnetic Flux Tubes in Dense QCD}}, \href{http://dx.doi.org/10.1103/PhysRevD.80.125007}{Phys. Rev. D {\bfseries 80} (2009) 125007} [\href{http://arxiv.org/abs/0907.1278}{{\arxivfont 0907.1278 [hep-ph]}}].

\bibitem{Eto:2009bh}
M.~Eto, E.~Nakano, and M.~Nitta, {\itshape {Effective world-sheet theory of color magnetic flux tubes in dense QCD}}, \href{http://dx.doi.org/10.1103/PhysRevD.80.125011}{Phys. Rev. D {\bfseries 80} (2009) 125011} [\href{http://arxiv.org/abs/0908.4470}{{\arxivfont 0908.4470 [hep-ph]}}].

\bibitem{Eto:2009tr}
M.~Eto, M.~Nitta, and N.~Yamamoto, {\itshape {Instabilities of Non-Abelian Vortices in Dense QCD}}, \href{http://dx.doi.org/10.1103/PhysRevLett.104.161601}{Phys. Rev. Lett. {\bfseries 104} (2010) 161601} [\href{http://arxiv.org/abs/0912.1352}{{\arxivfont 0912.1352 [hep-ph]}}].

\bibitem{Eto:2013hoa}
M.~Eto, Y.~Hirono, M.~Nitta, and S.~Yasui, {\itshape {Vortices and Other Topological Solitons in Dense Quark Matter}}, \href{http://dx.doi.org/10.1093/ptep/ptt095}{PTEP {\bfseries 2014} no.~1, (2014) 012D01} [\href{http://arxiv.org/abs/1308.1535}{{\arxivfont 1308.1535 [hep-ph]}}].

\bibitem{Mermin:1979zz}
N.~D. Mermin, {\itshape {The topological theory of defects in ordered media}}, \href{http://dx.doi.org/10.1103/RevModPhys.51.591}{Rev. Mod. Phys. {\bfseries 51} (1979) 591--648}.

\bibitem{Poenaru1977}
V.~Poenaru and G.~Toulouse, {\itshape {The crossing of defects in ordered media and the topology of 3-manifolds}}, \href{http://dx.doi.org/10.1051/jphys:01977003808088700}{J. Phys. (Paris) {\bfseries 38} (1977) 887}.

\bibitem{vol77}
G.~E. Volovik and V.~P. Mineev, {\itshape {Investigation of singularities in superfluid He3 and in liquid crystals by the homotopic topology methods}}, Zh. Eksp. Teor. Fiz {\bfseries 72} (1977) 2256. [Sov. Phys. JETP {\bf 45}, 1186 (1977)].

\bibitem{Lavrentovich2001}
O.~D. Lavrentovich and M.~Kleman, \href{http://dx.doi.org/10.1007/0-387-21642-1_5}{{\itshape Cholesteric {Liquid} {Crystals}: {Defects} and {Topology}},} in Chirality in {Liquid} {Crystals}, H.-S. Kitzerow and C.~Bahr, eds., pp.~115--158.
\newblock Springer New York, New York, NY, 2001.

\bibitem{Balachandran:1983pf}
A.~P. Balachandran, F.~Lizzi, and V.~G.~J. Rodgers, {\itshape {Topological Symmetry Breakdown in Cholesterics, Nematics, and $^{3}$He}}, \href{http://dx.doi.org/10.1103/PhysRevLett.52.1818}{Phys. Rev. Lett. {\bfseries 52} (1984) 1818--1821}.

\bibitem{salomaaRMP}
M.~M. Salomaa and G.~E. Volovik, {\itshape Quantized vortices in superfluid $^{3}\mathrm{He}$}, \href{http://dx.doi.org/10.1103/RevModPhys.59.533}{Rev. Mod. Phys. {\bfseries 59} (Jul, 1987) 533--613}.

\bibitem{Volovik:2003fe}
G.~E. Volovik, {\itshape {The Universe in a helium droplet}}, vol.~117.
\newblock 2006.

\bibitem{Semenoff:2006vv}
G.~W. Semenoff and F.~Zhou, {\itshape {Discrete symmetries and 1/3-quantum vortices in condensates of F=2 cold atoms}}, \href{http://dx.doi.org/10.1103/PhysRevLett.98.100401}{Phys. Rev. Lett. {\bfseries 98} (2007) 100401} [\href{http://arxiv.org/abs/cond-mat/0610162}{{\arxivfont cond-mat/0610162}}].

\bibitem{Kobayashi:2008pk}
M.~Kobayashi, Y.~Kawaguchi, M.~Nitta, and M.~Ueda, {\itshape {Collision Dynamics and Rung Formation of Non-Abelian Vortices}}, \href{http://dx.doi.org/10.1103/PhysRevLett.103.115301}{Phys. Rev. Lett. {\bfseries 103} (2009) 115301} [\href{http://arxiv.org/abs/0810.5441}{{\arxivfont 0810.5441 [cond-mat.other]}}].

\bibitem{Kawaguchi:2012ii}
Y.~Kawaguchi and M.~Ueda, {\itshape {Spinor Bose-Einstein condensates}}, \href{http://dx.doi.org/10.1016/j.physrep.2012.07.005}{Phys. Rept. {\bfseries 520} (2012) 253--381}.

\bibitem{Borgh:2016cco}
M.~O. Borgh and J.~Ruostekoski, {\itshape {Core Structure and Non-Abelian Reconnection of Defects in a Biaxial Nematic Spin-2 Bose-Einstein Condensate}}, \href{http://dx.doi.org/10.1103/PhysRevLett.117.275302}{Phys. Rev. Lett. {\bfseries 117} no.~27, (2016) 275302} [\href{http://arxiv.org/abs/1611.09735}{{\arxivfont 1611.09735 [cond-mat.quant-gas]}}]. [Erratum: Phys.Rev.Lett. 118, 129901 (2017)].

\bibitem{Mawson:2018klj}
T.~Mawson, T.~Petersen, J.~Slingerland, and T.~Simula, {\itshape {Braiding and Fusion of Non-Abelian Vortex Anyons}}, \href{http://dx.doi.org/10.1103/PhysRevLett.123.140404}{Phys. Rev. Lett. {\bfseries 123} no.~14, (2019) 140404} [\href{http://arxiv.org/abs/1805.10009}{{\arxivfont 1805.10009 [cond-mat.quant-gas]}}].

\bibitem{Annala:2022bdd}
T.~Annala, R.~Zamora-Zamora, and M.~M{\"o}tt{\"o}nen, {\itshape {Topologically protected vortex knots and links}}, \href{http://dx.doi.org/10.1038/s42005-022-01071-2}{Commun. Phys. {\bfseries 5} no.~1, (2022) 309}.

\bibitem{Rajamaki:2023ymv}
H.~Rajam{\"a}ki, T.~Annala, and M.~M{\"o}tt{\"o}nen, {\itshape {Topologically Protected Vortex Knots in an Experimentally Realizable System}}, \href{http://dx.doi.org/10.1103/PhysRevLett.133.236604}{Phys. Rev. Lett. {\bfseries 133} no.~23, (2024) 236604} [\href{http://arxiv.org/abs/2308.09825}{{\arxivfont 2308.09825 [cond-mat.quant-gas]}}].

\bibitem{Kobayashi:2024aip}
M.~Kobayashi, Y.~Nozaki, Y.~Koda, and M.~Nitta, {\itshape {Quantum Knots that Never Come Untied}}, \href{http://arxiv.org/abs/2410.07470}{{\arxivfont 2410.07470 [cond-mat.quant-gas]}}.

\bibitem{Masuda:2016vak}
K.~Masuda and M.~Nitta, {\itshape {Half-quantized non-Abelian vortices in neutron $^3P_2$ superfluids inside magnetars}}, \href{http://dx.doi.org/10.1093/ptep/ptz138}{PTEP {\bfseries 2020} no.~1, (2020) 013D01} [\href{http://arxiv.org/abs/1602.07050}{{\arxivfont 1602.07050 [nucl-th]}}].

\bibitem{Masaki:2021hmk}
Y.~Masaki, T.~Mizushima, and M.~Nitta, {\itshape {Non-Abelian half-quantum vortices in $^3$P$_2$ topological superfluids}}, \href{http://dx.doi.org/10.1103/PhysRevB.105.L220503}{Phys. Rev. B {\bfseries 105} no.~22, (2022) L220503} [\href{http://arxiv.org/abs/2107.02448}{{\arxivfont 2107.02448 [cond-mat.supr-con]}}].

\bibitem{Masaki:2023rtn}
Y.~Masaki, T.~Mizushima, and M.~Nitta, \href{http://dx.doi.org/10.1016/B978-0-323-90800-9.00225-0}{{\itshape {Non-Abelian Anyons and Non-Abelian Vortices in Topological Superconductors}},} in Encyclopedia of Condensed Matter Physics (Second Edition), vol.~2, pp.~755--794.
\newblock Elsevier, 2024.
\newblock \href{http://arxiv.org/abs/2301.11614}{{\arxivfont 2301.11614 [cond-mat.supr-con]}}.

\bibitem{Kobayashi:2022moc}
M.~Kobayashi and M.~Nitta, {\itshape {Core structures of vortices in Ginzburg-Landau theory for neutron $^3P_2$ superfluids}}, \href{http://dx.doi.org/10.1103/PhysRevC.105.035807}{Phys. Rev. C {\bfseries 105} no.~3, (2022) 035807} [\href{http://arxiv.org/abs/2203.09300}{{\arxivfont 2203.09300 [nucl-th]}}].

\bibitem{Kobayashi:2022dae}
M.~Kobayashi and M.~Nitta, {\itshape {Proximity effects of vortices in neutron $^3$P$_2$ superfluids in neutron stars: Vortex core transitions and covalent bonding of vortex molecules}}, \href{http://dx.doi.org/10.1103/PhysRevC.107.045801}{Phys. Rev. C {\bfseries 107} no.~4, (2023) 045801} [\href{http://arxiv.org/abs/2209.07205}{{\arxivfont 2209.07205 [nucl-th]}}].

\bibitem{Marmorini:2020zfp}
G.~Marmorini, S.~Yasui, and M.~Nitta, {\itshape {Pulsar glitches from quantum vortex networks}}, \href{http://dx.doi.org/10.1038/s41598-024-56383-w}{Sci. Rep. {\bfseries 14} no.~1, (2024) 7857} [\href{http://arxiv.org/abs/2010.09032}{{\arxivfont 2010.09032 [astro-ph.HE]}}].

\bibitem{Hattori:2026bvb}
T.~Hattori, M.~Nitta, and K.~Sekizawa, {\itshape {Formation of bound composite vortices of a singly-quantized $^1$S$_0$ vortex and half-quantized $^3$P$_2$ vortices in the $^1$S$_0$-$^3$P$_2$ coexisting phase in neutron stars}}, \href{http://arxiv.org/abs/2605.28718}{{\arxivfont 2605.28718 [nucl-th]}}.

\bibitem{Horvathy:1985jr}
P.~A. Horvathy, {\itshape {The Nonabelian {Aharonov-Bohm} Effect}}, \href{http://dx.doi.org/10.1103/PhysRevD.33.407}{Phys. Rev. D {\bfseries 33} (1986) 407--414}.

\bibitem{Sundrum:1986ub}
R.~Sundrum and L.~J. Tassie, {\itshape {Nonabelian {Aharonov-Bohm} Effects, Feynman Paths, and Topology}}, \href{http://dx.doi.org/10.1063/1.527067}{J. Math. Phys. {\bfseries 27} (1986) 1566--1570}.

\bibitem{Alford:1989ch}
M.~G. Alford, J.~March-Russell, and F.~Wilczek, {\itshape {Discrete Quantum Hair on Black Holes and the Nonabelian {Aharonov-Bohm} Effect}}, \href{http://dx.doi.org/10.1016/0550-3213(90)90512-C}{Nucl. Phys. B {\bfseries 337} (1990) 695--708}.

\bibitem{Preskill:1990bm}
J.~Preskill and L.~M. Krauss, {\itshape {Local Discrete Symmetry and Quantum Mechanical Hair}}, \href{http://dx.doi.org/10.1016/0550-3213(90)90262-C}{Nucl. Phys. B {\bfseries 341} (1990) 50--100}.

\bibitem{Bais:1991pe}
F.~A. Bais, P.~van Driel, and M.~de~Wild~Propitius, {\itshape {Quantum symmetries in discrete gauge theories}}, \href{http://dx.doi.org/10.1016/0370-2693(92)90773-W}{Phys. Lett. B {\bfseries 280} (1992) 63--70} [\href{http://arxiv.org/abs/hep-th/9203046}{{\arxivfont hep-th/9203046}}].

\bibitem{Bais:1992ca}
F.~A. Bais, P.~van Driel, and M.~de~Wild~Propitius, {\itshape {Anyons in discrete gauge theories with Chern-Simons terms}}, \href{http://dx.doi.org/10.1016/0550-3213(93)90073-X}{Nucl. Phys. B {\bfseries 393} (1993) 547--570} [\href{http://arxiv.org/abs/hep-th/9203047}{{\arxivfont hep-th/9203047}}].

\bibitem{Krauss:1988zc}
L.~M. Krauss and F.~Wilczek, {\itshape {Discrete Gauge Symmetry in Continuum Theories}}, \href{http://dx.doi.org/10.1103/PhysRevLett.62.1221}{Phys. Rev. Lett. {\bfseries 62} (1989) 1221}.

\bibitem{Altarelli:2010gt}
G.~Altarelli and F.~Feruglio, {\itshape {Discrete Flavor Symmetries and Models of Neutrino Mixing}}, \href{http://dx.doi.org/10.1103/RevModPhys.82.2701}{Rev. Mod. Phys. {\bfseries 82} (2010) 2701--2729} [\href{http://arxiv.org/abs/1002.0211}{{\arxivfont 1002.0211 [hep-ph]}}].

\bibitem{Feruglio:2017spp}
F.~Feruglio, {\itshape {Are neutrino masses modular forms?}}, \href{http://dx.doi.org/10.1142/9789813238053_0012}{pp.~227--266}.
\newblock 2019.
\newblock \href{http://arxiv.org/abs/1706.08749}{{\arxivfont 1706.08749 [hep-ph]}}.

\bibitem{Ishimori:2010au}
H.~Ishimori, T.~Kobayashi, H.~Ohki, Y.~Shimizu, H.~Okada, and M.~Tanimoto, {\itshape {Non-Abelian Discrete Symmetries in Particle Physics}}, \href{http://dx.doi.org/10.1143/PTPS.183.1}{Prog. Theor. Phys. Suppl. {\bfseries 183} (2010) 1--163} [\href{http://arxiv.org/abs/1003.3552}{{\arxivfont 1003.3552 [hep-th]}}].

\bibitem{Ovrut:1977cn}
B.~A. Ovrut, {\itshape {Isotropy Subgroups of SO(3) and Higgs Potentials}}, \href{http://dx.doi.org/10.1063/1.523660}{J. Math. Phys. {\bfseries 19} (1978) 418}.

\bibitem{Etesi:1996urw}
G.~Etesi, {\itshape {Spontaneous symmetry breaking in SO(3) gauge theory to discrete subgroups}}, \href{http://dx.doi.org/10.1063/1.531470}{J. Math. Phys. {\bfseries 37} (1996) 1596--1602} [\href{http://arxiv.org/abs/hep-th/9706029}{{\arxivfont hep-th/9706029}}].

\bibitem{King:2018fke}
S.~F. King and Y.-L. Zhou, {\itshape {Spontaneous breaking of $SO(3)$ to finite family symmetries with supersymmetry - an $A_4$ model}}, \href{http://dx.doi.org/10.1007/JHEP11(2018)173}{JHEP {\bfseries 11} (2018) 173} [\href{http://arxiv.org/abs/1809.10292}{{\arxivfont 1809.10292 [hep-ph]}}].

\bibitem{Gaiotto:2014kfa}
D.~Gaiotto, A.~Kapustin, N.~Seiberg, and B.~Willett, {\itshape {Generalized Global Symmetries}}, \href{http://dx.doi.org/10.1007/JHEP02(2015)172}{JHEP {\bfseries 02} (2015) 172} [\href{http://arxiv.org/abs/1412.5148}{{\arxivfont 1412.5148 [hep-th]}}].

\bibitem{Gukov:2008sn}
S.~Gukov and E.~Witten, {\itshape {Rigid Surface Operators}}, \href{http://dx.doi.org/10.4310/ATMP.2010.v14.n1.a3}{Adv. Theor. Math. Phys. {\bfseries 14} no.~1, (2010) 87--178} [\href{http://arxiv.org/abs/0804.1561}{{\arxivfont 0804.1561 [hep-th]}}].

\bibitem{Heidenreich:2021xpr}
B.~Heidenreich, J.~McNamara, M.~Montero, M.~Reece, T.~Rudelius, and I.~Valenzuela, {\itshape {Non-invertible global symmetries and completeness of the spectrum}}, \href{http://dx.doi.org/10.1007/JHEP09(2021)203}{JHEP {\bfseries 09} (2021) 203} [\href{http://arxiv.org/abs/2104.07036}{{\arxivfont 2104.07036 [hep-th]}}].

\bibitem{Kobayashi:2011xb}
S.~Kobayashi, M.~Kobayashi, Y.~Kawaguchi, M.~Nitta, and M.~Ueda, {\itshape {Abe homotopy classification of topological excitations under the topological influence of vortices}}, \href{http://dx.doi.org/10.1016/j.nuclphysb.2011.11.003}{Nucl. Phys. B {\bfseries 856} (2012) 577--606} [\href{http://arxiv.org/abs/1110.1478}{{\arxivfont 1110.1478 [math-ph]}}].

\bibitem{Chatterjee:2017jsi}
C.~Chatterjee and M.~Nitta, {\itshape {BPS Alice strings}}, \href{http://dx.doi.org/10.1007/JHEP09(2017)046}{JHEP {\bfseries 09} (2017) 046} [\href{http://arxiv.org/abs/1703.08971}{{\arxivfont 1703.08971 [hep-th]}}].

\bibitem{Chatterjee:2017hya}
C.~Chatterjee and M.~Nitta, {\itshape {The effective action of a BPS Alice string}}, \href{http://dx.doi.org/10.1140/epjc/s10052-017-5352-1}{Eur. Phys. J. C {\bfseries 77} no.~11, (2017) 809} [\href{http://arxiv.org/abs/1706.10212}{{\arxivfont 1706.10212 [hep-th]}}].

\bibitem{Leonhardt:2000km}
U.~Leonhardt and G.~E. Volovik, {\itshape {How to create Alice string (half quantum vortex) in a vector Bose-Einstein condensate}}, \href{http://dx.doi.org/10.1134/1.1312008}{Pisma Zh. Eksp. Teor. Fiz. {\bfseries 72} (2000) 66--70} [\href{http://arxiv.org/abs/cond-mat/0003428}{{\arxivfont cond-mat/0003428}}].

\bibitem{Sato:2018nqy}
R.~Sato, F.~Takahashi, and M.~Yamada, {\itshape {Unified Origin of Axion and Monopole Dark Matter, and Solution to the Domain-wall Problem}}, \href{http://dx.doi.org/10.1103/PhysRevD.98.043535}{Phys. Rev. D {\bfseries 98} no.~4, (2018) 043535} [\href{http://arxiv.org/abs/1805.10533}{{\arxivfont 1805.10533 [hep-ph]}}].

\bibitem{Chatterjee:2019rch}
C.~Chatterjee, T.~Higaki, and M.~Nitta, {\itshape {Note on a solution to domain wall problem with the Lazarides-Shafi mechanism in axion dark matter models}}, \href{http://dx.doi.org/10.1103/PhysRevD.101.075026}{Phys. Rev. D {\bfseries 101} no.~7, (2020) 075026} [\href{http://arxiv.org/abs/1903.11753}{{\arxivfont 1903.11753 [hep-ph]}}].

\bibitem{Schwarz:1982ec}
A.~S. Schwarz, {\itshape Field theories with no local conservation of the electric charge}, \href{http://dx.doi.org/10.1016/0550-3213(82)90190-0}{Nucl. Phys. B {\bfseries 208} (1982) 141--158}.

\bibitem{Hidaka:2024kfx}
Y.~Hidaka, M.~Nitta, and R.~Yokokura, {\itshape {Selection rules of topological solitons from non-invertible symmetries in axion electrodynamics}}, \href{http://dx.doi.org/10.1007/JHEP09(2025)082}{JHEP {\bfseries 09} (2025) 082} [\href{http://arxiv.org/abs/2411.05434}{{\arxivfont 2411.05434 [hep-th]}}].

\end{thebibliography}
\endgroup

\end{document}